\documentclass{aa} 
\usepackage{graphicx}
\usepackage{txfonts}
 \usepackage[T1]{fontenc}
\usepackage{color}
\usepackage[]{natbib}
\usepackage[switch]{lineno}

        \newcommand{\Log}{${\rm Log}$}
        
        \newcommand{\msun}{\ensuremath{\mathrm{M}_{\odot}}}
        
        \newcommand{\sfr}{M$_{\odot}$ yr$^{-1}$}

\begin{document}

   \title{ALMA constraints on the faint millimetre source
number counts and their contribution to the cosmic infrared background}

   \author{S. Carniani
          \inst{1,2,3},
          R. Maiolino
          \inst{2,3},         
          G. De Zotti
          \inst{4,5},
          M. Negrello
          \inst{5},
          A. Marconi \inst{1},
          M. S. Bothwell
          \inst{2},
          P. Capak 
          \inst{6,7},
          C. Carilli  \inst{2,8},
          M. Castellano \inst{9},
          S. Cristiani \inst{10},
          A. Ferrara \inst{11},
          A. Fontana \inst{9},
          S. Gallerani \inst{11},
          G. Jones \inst{12},
          K. Ohta \inst{13},
          K. Ota        \inst{2,3},
          L. Pentericci \inst{9},
          P. Santini \inst{9},
          K. Sheth \inst{6,7},
          L. Vallini \inst{14},
          E. Vanzella \inst{15},
          J. Wagg \inst{2,3,16},
          R. J. Williams \inst{2,3}
          }

   \institute{Dipartimento di Fisica e Astronomia, Universit\`a di Firenze, Via G. Sansone 1, I-50019, Sesto Fiorentino (Firenze), Italy
                      \and
        Cavendish Laboratory, University of Cambridge, 19 J. J. Thomson Ave., Cambridge CB3 0HE, UK 
                \and
        Kavli Institute for Cosmology, University of Cambridge, Madingley Road, Cambridge CB3 0HA, UK
         \and
         SISSA, Via Bonomea 265, 34136, Trieste, Italy 
         \and
         INAF-Osservatorio Astronomico di Padova, Vicolo dell' Osservatorio 5, I-35122 Padova, Italy 
         \and
         Spitzer Science Center, 314-6 Caltech, 1201 East California Boulevard, Pasadena, CA 91125, USA 
         \and 
         Department of Astronomy, 249-17 Caltech, 1201 East California Boulevard, Pasadena, CA 91125, USA 
         \and
         National Radio Astronomy Observatory, P.O. Box 0, Socorro, NM 87801, USA 
         \and
         INAF - Osservatorio Astronomico di Roma, via Frascati 33, 00040 Monteporzio, Italy 
         \and 
         INAF- Trieste Astronomical Observatory, via Tiepolo 11, 34143 Trieste, Italy 
         \and
         Scuola Normale Superiore, Piazza dei Cavalieri 7, 56126, Pisa, Italy 
         \and
         Physics Department, New Mexico Tech, Socorro, NM 87801, USA 
         \and
         Department of Astronomy, Kyoto University, Kyoto 606-8502, Japan
         \and
                Dipartimento di Fisica e Astronomia, Universita' di Bologna, Viale Berti Pichat 6/2, I-30127, Bologna, Italy  
\and
INAF-Osservatorio Astronomico di Bologna, Via Ranzani 1, I-40127 Bologna, Italy 
\and
Square Kilometre Array Organization, Jodrell Bank Observatory, Lower Withington, Macclesfield, Cheshire SK11 9DL, UK 
}


\abstract{We have analysed 18 ALMA continuum maps in Bands 6 and 7, with rms down to 7.8 $\mu$Jy,
to  derive differential number counts down to 60 $\mu$Jy and 100
$\mu$Jy at $\lambda =$1.3 mm and $\lambda =$1.1 mm, respectively.
Furthermore, the  non-detection of faint
sources in the deepest ALMA field enabled us to set tight upper limits on the number counts down to 30 $\mu$Jy. This is a factor of four deeper than the currently most stringent upper limit.
The area covered by the combined fields  is $\rm 9.5\times 10^{-4}~deg^2$ at 1.1~mm
and $\rm 6.6\times 10^{-4}~deg^{2}$ at 1.3mm.
With respect to previous works,
we improved the source extraction method by  requiring that the dimension of the detected sources be consistent
with the beam size. This method enabled us to remove spurious detections that have plagued
the purity of the catalogues in previous studies.
 We detected 50 faint sources (at fluxes $<$ 1 mJy) with signal-to-noise (S/N) $>$ 3.5 down to 60 $\mu$Jy, hence improving
the statistics by a factor of four relative to previous studies. 
The inferred differential number counts are $\rm dN/d(Log_{10}S)=1\times10^5~deg^2$ 
at a 1.1~mm flux $\rm S_{\lambda = 1.1~mm} = 130~\mu Jy$, 
and $\rm dN/d(Log_{10}S)=1.1\times10^5~deg^2$ at  a 1.3~mm flux $\rm S_{\lambda = 1.3~mm} = 60~\mu Jy$.
At the faintest flux limits probed by our data, i.e. 30 $\mu$Jy and 40 $\mu$Jy, we obtain  upper limits on
the differential number counts of $\rm  dN/d(Log_{10}S) < 7\times10^5~deg^2$ and $\rm  dN/d(Log_{10}S) < 3\times10^5~deg^2$,
respectively.
Determining the fraction of cosmic infrared background (CIB) resolved by the ALMA observations was
hampered by the large uncertainties plaguing the CIB measurements (a factor of four in
flux). However,
our results provide a  new lower limit to CIB intensity  of 17.2 ${\rm Jy \ deg^{-2}}$ at 1.1 mm  and of  12.9 ${\rm Jy \ deg^{-2}}$ at 1.3 mm. Moreover, the
flattening of the integrated number counts at faint fluxes strongly suggests
that  we are probably close to the CIB intensity. Our data imply that
galaxies with star formation rate (SFR)$ < 40 ~M_{\odot}/yr$ certainly contribute
less than 50\% to the CIB (and probably a much lower percentage)  while
more than 50\% of the CIB must be produced by galaxies with $\rm SFR > 40 ~M_{\odot}/yr$.
The differential number counts are in nice agreement 
with recent semi-analytical models of galaxy formation even as low as our faint
fluxes. Consequently, this supports the galaxy evolutionary scenarios and assumptions
made in these models.}

\authorrunning{Carniani et al.}
\titlerunning{ALMA constraints on the faint millimetre source
number counts and their contribution to the CIB}
   \maketitle

\section{Introduction}
The extragalactic background light (EBL) is a diffuse and isotropic radiation in the Universe, covering the range between ultraviolet (UV) and far
infrared (FIR) wavelengths \citep{Fixsen:1998}. After the CMB, the EBL represents the second most energetic background.   The IR/mm spectrum of
the EBL was first estimated by \cite{Puget:1996} using data from Far Infrared Absolute Spectrometer on the Cosmic Background Explorer (COBE) satellite.
The EBL spectral energy
  distribution is composed of two peaks: the cosmic optical background
  (COB) and the cosmic infrared background (CIB). The former is caused by
  the radiation from stars, while the latter is due to UV/optical light absorbed by
  dust and reradiated in the infrared wavelength range.  By measuring the integrated flux of the 
  two components, the ratio between the COB and CIB is of the order of unity \citep{Dole:2006}, which suggests that half of the star light emission is absorbed by dust in galaxies. Therefore, the EBL
contains information about star formation processes and galaxy
evolution in the Universe. 
The study of this emission helps us to understand the star formation evolution throughout the history of the Universe. 

 The mixture of source populations 
contributing to the CIB depends strongly on the specific wavelength (e.g. \citealt{Viero:2013a,Cai:2013}).
At millimetre wavelengths ($\lambda \sim 1.1-1.3~\mu m$), a significant
percentage ($\sim 30$\%) of the CIB is emitted by submillimetre
galaxies (e.g. \citealt{Viero:2013a,Cai:2013}). The definition of submillimetre
galaxies (SMGs) is somewhat loose and  is generally meant to
indicate the bright end of the population of sources emitting at submm
wavelengths, initially discovered by bolometers on single-dish
telescopes. SMGs are predominately high-redshift ($z \ge 1$)
star-forming galaxies with star formation rates (SFRs) approaching
1000 \sfr, or even higher (e.g. \citealt{Blain:2002}). In these galaxies, the
bulk of the UV/optical emission from young
stars is absorbed by the surrounding dust, which is re-emitted at
FIR wavelengths \citep{Casey:2014a}.
Recent studies have shown that massive red-and-dead galaxies share
the same clustering properties as SMGs.  Thus,  SMGs
may be the progenitors of local massive elliptical galaxies
\citep{simpson:2014,Toft:2014}. 


However, SMGs represent an extreme class of objects, not
representative of the bulk of the galaxy population at high $z$ ($z\gtrsim1$). 
Most high-$z$ galaxies show a 
much lower SFR and are likely associated with systems that evolve through
secular processes (e.g. \citealt{Rodighiero:2011}). For the bulk of the
high-$z$ population, the rest-frame far-IR/submm properties (which
provide information on obscured star formation and dust content), are still poorly known at
mm/submm fluxes below 1 mJy.


In the past decades the SCUBA and LABOCA  single-dish surveys  resolved 20$\%$ to
40$\%$ of the CIB at 850 $\mu$m (e.g.
\citealt{Eales:1999,Coppin:2006,Weis:2009}) and 10$\%$ to 20$\%$ of the
CIB at 1.1 mm  with deep single-dish surveys using the AzTEC camera (e.g. \citealt{Scott:2010}).
Until recently, the number counts of fainter sources ($S<1$~mJy) were
not well constrained because of the limited sensitivity.  However, the
observation of lensed galaxies, hence reaching somewhat deeper flux limits (e.g.
 \citealt{Cowie:2002, Knudsen:2008, Johansson:2011,
Chen:2013,Chen:2013a}) suggests that more than 50$\%$ of the CIB is
emitted by faint sources with flux densities $<2$ mJy.

Recently, the  number counts of faint mm sources, at fluxes fainter than 1 mJy, have been inferred thanks to high-sensitivity and high-resolution observations obtained with the Atacama Large Millimeter/submillimeter Array (ALMA). \cite{Hatsukade:2013} claim to have
resolved $\sim80\%$ ( $\sim$ 13 Jy deg$^{-2}$) of the CIB at 1.3~mm  exploring faint (0.1 - 1 mJy) sources with signal-to-noise (S/N)  $\geq4$ extracted from  ALMA
data. A similar result has been obtained at 1.2  mm by \cite{Ono:2014}, who
suggest
that the main contribution to the CIB comes from faint star-forming galaxies
with SFR $< 30 \ {\rm M}_\odot /$yr. 
However, as we  discuss later, the uncertainties on the CIB spectrum are large, and once these are taken into account, the fraction of
resolved background at these fluxes, as well as
the identification of the sources
contributing to the bulk of the CIB, are much more uncertain than is given in these
papers.  Source number counts with deep millimetre 
observations can provide a tight lower limit on the CIB intensity.
Moreover, the slope of the faint counts constrains contributions to the CIB
from still fainter sources.
Finally, the detected faint millimetre sources  can be  targets for  future spectroscopic  observations aimed at
understanding the properties of this faint population,  which is more
representative of the bulk of the galaxy population than past (bright)
millimeter sources.

We  used imaging from ALMA with a sensitivity down to 7.8
$\mu$Jy/beam (rms), which enabled us to achieve some of the faintest
continuum detections at 1.1 mm and 1.3 mm with flux densities
down to 60~$\mu$Jy. The source counts presented here thus provide
constraints on models of galaxy evolution and predictions for
future ALMA follow-up surveys.

This paper is organised as follows. In Section 2, we describe the ALMA
observations used in this work. Section 3 is focused on the source
extraction technique. Section 4 presents the number counts we derived
and the comparison between  our results and recent galaxy
formation models. In Section 5, we discuss and summarise our results. 


\section{Observations and data reduction}\label{sec:data}

Our source extraction is applied to 18 continuum maps with high
sensitivities, which were obtained in ALMA Cycle 0 and Cycle 1.  For
our analysis we focused on observations in Bands 6 and 7,
since these are some of the deepest observations available.  In
  this section, we  describe in detail the ALMA Bands 6 and 7 data sets
  used for the analysis.
        
The faintest sources are detected in three ALMA data sets taken by
\cite{Maiolino:2015}, who targeted three Ly$\alpha$ emitters at
z$\sim$6-7: BDF-3299, BDF-521 and SDF-46975 \citep{Vanzella:2011, Ono:2012}.
The ${\rm BDF\textrm{-}3299}$ data (Field \emph{a} in
Table \ref{tab:field}) were observed during two different epochs: a
first observation between October and November 2013 and a second one in April
2014. The target was observed with 27 12m antennae array in 2013 and 36
12m antennae array in 2014 with a maximum baseline of 1270 m. The flux
densities were  calibrated with the observation data of J223-3137 and J2247-3657.
The total on-source integration time was $\sim$5.2 hr. The other two
sources, BDF-521 and ${\rm SDF\textrm{-}46975}$, were observed in
November 2013 and March 2014, respectively, (Fields \emph{c} and
\emph{e} in Table \ref{tab:field}).  In the
extended configuration of 17-1284 m baseline for BDF-521, 29 12m antennae were used  and 40 12m
antennae with a maximum baseline of 422 m for SDF-46975. We used the 
observations of J223-3137 to  calibrate the flux density. The total on
source observing time was about 83 min and 121 min, respectively, for
the two targets.
   
We analysed nine continuum maps (fields \emph{j}-\emph{r} in
Table \ref{tab:field}) taken by PI Capak, who targeted [CII] 
emission line from sources at high redshift ($z\geq5$). The data were taken 
in November 2013 using 20 antennae in band 7. The total on-source
integration was about ~20 min for each.

We also used ALMA data (field \emph{b} in Table \ref{tab:field}) 
  for the Ly$\alpha$ emitter at z = 7.215, SXDF-NB1006-2 
  (PI K. Ota; \citealt{Shibuya:2012a}). The target was observed on
   May 3-4, 2014 with a maximum baseline of $\sim$ 558 m. The total on-source observing time of the 37 12 m antennae was 106 min. The flux
densities were scaled with the observation data of J0215-0222.

In addition to these maps, we analysed public archival ALMA data to
increase the number of detections at intermediate flux densities.  We
selected only continuum maps in Bands 6 and 7 with sensitivity $\leq~50\ \mu$Jy/beam
since we were interested in analysing the number count at flux
densities $< 1$ mJy that contribute to $> 60\%$ of the CIB
\citep{Ono:2014}. Therefore, we analysed the data with the highest
sensitivity taken by \cite{Willott:2013} (Fields \emph{f} and \emph{i}
in Table \ref{tab:field}), \cite{Ota:2014} (Field \emph{g} in
Table \ref{tab:field}), \cite{MacGregor:2013} (Field \emph{h} in
Table \ref{tab:field}), and \cite{Ouchi:2013} (Field \emph{d} in
Table \ref{tab:field}). Further details on the ALMA observations are
summarised in those papers.

All ALMA data were reduced using the CASA v4.2.1 package. 
The typical flux uncertainties 
in the millimeter regime are $\sim10\%$. 
The
continuum maps were extracted using all the line-free channels of the
four spectral windows. Unfortunately, in this version of CASA, the
data weights were not set proportionally to the channel width and
integration time, so they had to be adjusted whenever a continuum
image was made from spectral windows that did not have the same
channel width and index number. Furthermore, the data weights had to
be fixed when a dataset was composed of different observations taken at
different epochs with different integration times and different observing
(water vapour) conditions. In the case of
BDF-3299, BDF-521 and SDF-46975, the data weights were manually
re-scaled as a function of integration time and channel width. The
continuum maps were cleaned using the CASA task clean with
WEIGHTING = ``natural'', achieving a sensitivity  in the range between 7.8
$\mu$Jy/beam (which is the deepest ALMA observation at this wavelength and three times
deeper than data used in previous studies) and
52.1
$\mu$Jy/beam.  The correlator of each observation was
configured to provide four independent spectral windows, so the central
frequency $\nu_{\rm obs}$ in Table \ref{tab:field} is equivalent to the mean
frequency of the four bands.  The continuum map sensitivity and the area mapped in
each observation are summarised in Table \ref{tab:field}. The source extraction was
performed as far out as two primary beams, after masking the targeted source of
each observation, so as  not to bias the final counts determination. Around all of these sources we placed a 1\arcsec\ diameter
  mask ($\sim$ ALMA beam), since most of the main targets were non-spatially resolved. In the particular case where the main target is
  extended (e.g. \citealt{MacGregor:2013}), the dimension of the mask
  is as large as the size of the target, where the size of the target
  is estimated from its surface brightness emission down to
  $3\sigma$. In the worst case, we masked about 5$\%$ of the field of
  view. 
   The combined fields result in a total area of ${\rm \sim9.5\times10^{-4} 
\ deg^2}$  at 1.1 mm and ${\rm \sim6.6\times10^{-4} \  deg^2}$ at 1.3 mm
(which, in general, is  two times  larger than previous studies).

\begin{table*}
\caption{ALMA survey fields used in this paper, sorted by sensitivity.}           
\label{tab:field}      
\centering          
\begin{tabular}{c c c c c c}    
\\
Project         &       $\nu_{\rm obs}$  &      $\sigma$                        &         $\lambda$       &  Area & Field   \\
code            &               [GHz]                    &  [$\mu$Jy beam$^{-1}$]   &     [mm]            &       [$10^{-4}$ deg$^2$] &      \\
  & (1) & (2) & (3) & (4) & \\
\hline
2012.A.00040.S  &       230             &       7.8             &       1.28    &       1.17  & a    \\  
2012.1.00374.S          &      225      &     14.5    &   1.31    &   1.17  & b   \\ 
2012.1.00719.S  &       230             &       17.7    &       1.30    &       1.17  & c    \\ 
2011.1.00115.S          &       260             &       18.6    &       1.16    &       0.87  & d    \\ %
2012.1.00719.S  &       244             &       19.5    &       1.23    &       1.06  & e    \\ 
2011.1.00243.S  &       250             &       20.9    &       1.2             &       0.97  & f    \\ %
2011.0.00767.S          &       230             &       20.9    &       1.30    &       1.17  & g    \\
2012.1.00142.S  &       230             &       26.3    &       1.28    &       1.07  & h    \\ 
2011.1.00243.S  &       249             &       28.9    &       1.2             &       0.97  & i    \\ %
2012.1.00523.S  &       286             &       29.9    &       1.05    &       0.72   & j   \\ 
2012.1.00523.S  &       295             &       30.0    &       1.02    &       0.71   & k   \\
2012.1.00523.S  &       286             &       30.3    &       1.05    &       0.72   & l   \\
2012.1.00523.S  &       289             &       33.1    &       1.05    &       0.72   & m   \\
2012.1.00523.S  &       290             &       36.3    &       1.03    &       0.71   & n   \\
2012.1.00523.S  &       289             &       38.1    &       1.05    &       0.72   & o   \\
2012.1.00523.S  &       289             &       41.8    &       1.05    &       0.72   & p   \\
2012.1.00523.S  &       292             &       49.1    &       1.02    &       0.71   & q   \\
2012.1.00523.S  &       292             &       52.1    &       1.02    &       0.71   & r   \\ 

 \hline 
\end{tabular}

\tablefoot{ (1) Frequency in the observed frame corresponding to the mean frequency of the four  ALMA spectral window. (2) rms measured in each continuum map before primary beam correction. (3) Central wavelength in the observed frame. (4) Area in two primary beams. The size of the primary beam scales linearly with wavelength. }
 
 
\end{table*}

\section{Source extraction}

In total
we analysed 18 ALMA continuum maps to derive the number counts of sources at millimeter wavelengths. 
Since we do not yet know either the spectral energy distributions (SEDs) or the redshifts of our faint mm sources, we estimated the number counts at two different wavelengths, 1.1 mm and 1.3 mm, to minimise the effects of
wavelength extrapolation. The flux densities, $S$, of sources
detected at wavelengths less than 1.2 mm were scaled to the 1.1 mm flux density,  while  counts at $\lambda$$>1.2$ mm were scaled to 1.3 mm  using
a modified blackbody with values typical of SMGs at z=2. We adopted a spectral index $\beta = 2.0$ and dust temperature T = 35 K from \cite{Greve:2012}, who measured the SED from a sample of SMGs at the same wavelength range of our data.   
 As these  SED properties can be different for each source, in Appendix \ref{sec:appendix_A} we estimate the errors induced by varying  these parameters. 
In  the following, we describe the source extraction  method and the statistical assessment in detail.

\subsection{Source catalogue}\label{sec:source_extraction}
The source extraction was performed within an area as large as  two primary beams 
that has a diameter
of about 20'', before correcting for the primary beam attenuation.
We first extracted the sources fulfilling the following requirement: 1)
the source should be above the 3$\sigma$ threshold in its  continuum map (we  then
took a more conservative threshold of 3.5$\sigma$, as discussed later); 2)
the size of the 2D Gaussian fitting the putative source must be consistent, within the errors, with the beam size of the selected map (or at most marginally
resolved, within 1.5 times the beam size). Indeed, most 
faint sources are not expected to be spatially resolved at the resolution of our maps. Detections with dimensions smaller than the beam
must be associated with noise fluctuation of individual antennae or a group of antennae, or be caused by sidelobes of
bright sources.This additional source detection criterion enables us to greatly reduce (by a factor of 3)
the number of spurious sources, hence making the final catalogue much less prone to false detections than previous studies.

  \begin{figure}
   \centering
\includegraphics[width =.5\textwidth]{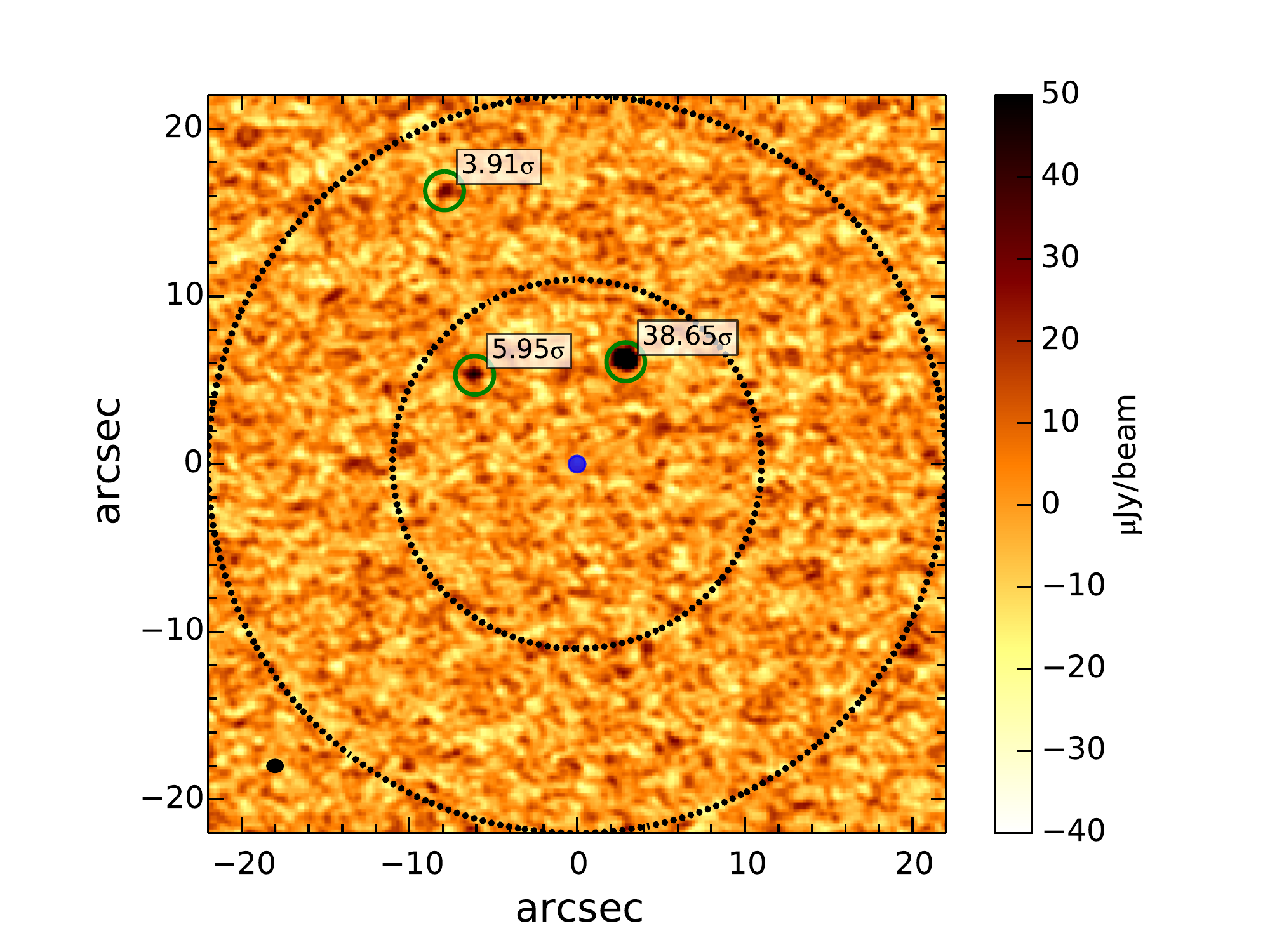}

\caption{Example of a Band 6 continuum map obtained with ALMA (Field
  $a$ Table \ref{tab:field}). The green circles represent the sources detected
  with S/N $>3.5\sigma$ that also fulfil the requirement of having a size
  consistent with the beam (or marginally resolved). The inner black dotted circle indicates the
  primary beam and the outer circle shows twice the primary beam.  The blue
  filled circle shows the masked region.
   The synthesised beam is indicated by a filled
  black ellipse in the lower left corner of the plot.}
 \label{fig:example}
   \end{figure}

\begin{figure}
   \centering 
   \includegraphics[width=.46\textwidth]{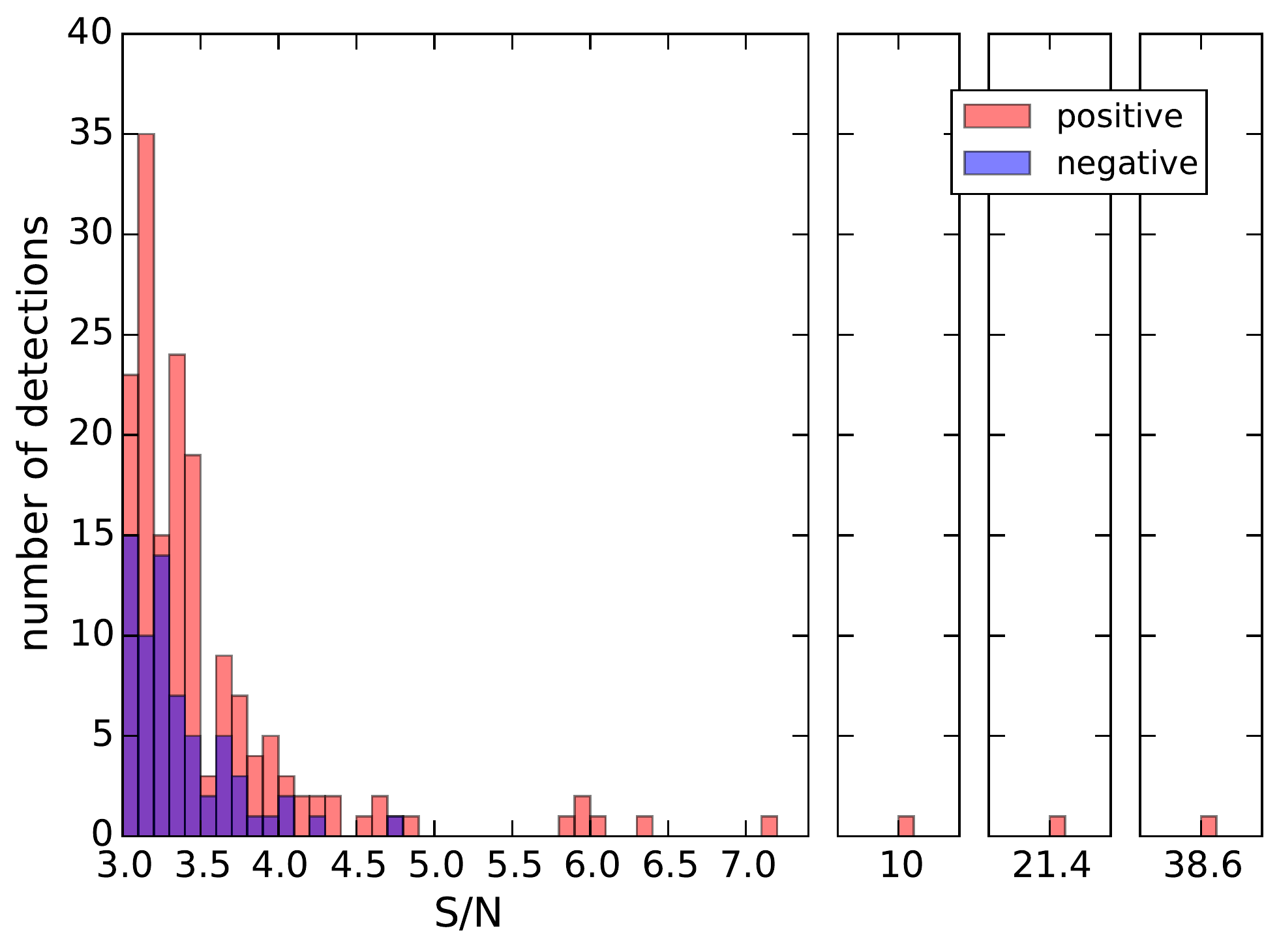}
\includegraphics[width
     =.48\textwidth]{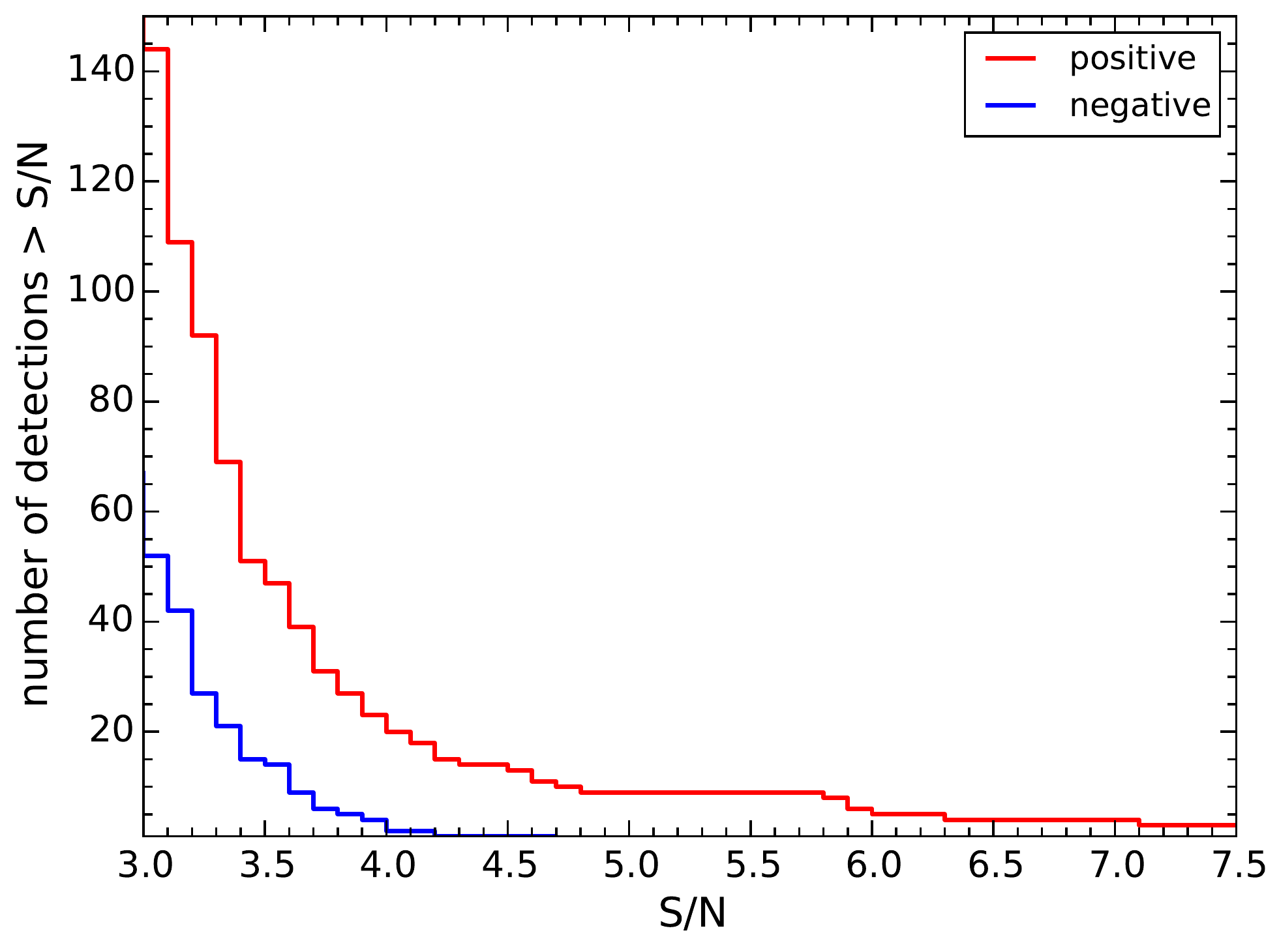}

\caption{\emph{Top:} number of positive (red) and negative (blue) sources detected in the 18 continuum maps, as a function of S/N. \emph{Bottom:} cumulative distribution of positive (red) and negative (blue) detections.}
 \label{fig:histo}
   \end{figure}

Figure \ref{fig:example} shows an example of an ALMA map in which the
source extraction was performed down to 3$\sigma$ with the above
requirements.  At this low significance level ($>3\sigma$), some of
these source candidates are likely to be spurious because of  noise
fluctuations.  To define a more solid detection threshold, we
estimated the number of spurious sources expected in the maps by
applying the source extraction method to the continuum maps, which were multiplied
by ${-1}$   to estimate the number of negative sources as a
  function of the S/N. Figure \ref{fig:histo}
shows the number of positive and negative sources as a function of
S/N. The number of the negative sources is almost always less than
that of the positive at S/N~$>3$, which suggests that down to
3$\sigma,$ some fraction of the positive detections are real (most of
the negative sources and false positive sources are removed from the
second requirement in the source extraction process).  It is also
clear because the
cumulative number of positive
sources is larger than that of the negative ones down to S/N =3. Moreover, 
simulations of blank fields, with exactly the same observing conditions as our
data (Appendix \ref{sec:appendix_A}), show that the number of positive 
and negative sources due to noise fluctuations are equal for any S/N level.  Because the
  number of positive sources was found to be larger than the number of
  negative ones at S/N $> 3.5 $ in
  previous works from \cite{Hatsukade:2013} and \cite{Ono:2014}, we decided to be conservative by
  only including  those objects that satisfy the S/N $> 3.5$ criterion in our catalogue  .



In the 18 continuum maps, we detected 50 sources with S/N in the range 3.5-38.4, and none of them appear to be marginally
resolved.  These statistics are a factor of four higher than previous studies \citep{Ono:2014}.

\subsection{Completeness and survey area}

\begin{figure}
   \centering
\includegraphics[width =.5\textwidth]{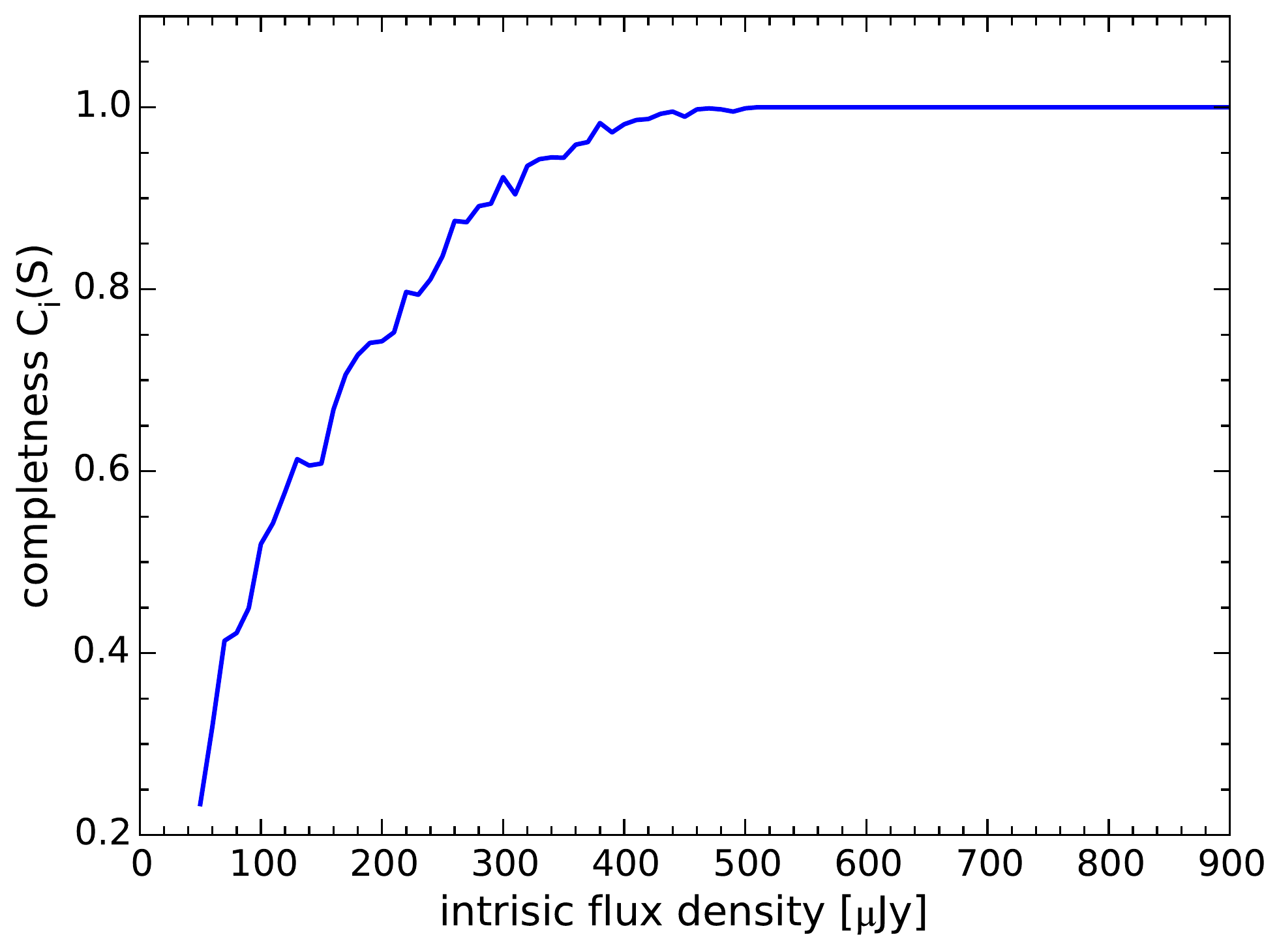}

\caption{Completeness $C_i(S)$ as a function of the flux density $S,$
  estimated from simulations. The solid curve is the result for field
  $a$ with rms = 7.8 $\mu$Jy/beam.}
 \label{fig:completeness}
   \end{figure}

For each ALMA map $i$, we estimated the completeness, $C_{i}(S)$, which
is the expected probability at which a real source with flux
$S$ can be detected within the entire field of view that we considered (i.e. two
primary beams). The $C_{i}(S)$ is calculated in each ALMA map
corrected for primary beam attenuation. To estimate the completeness
we inserted artificial sources with a given flux density $S$ (at
which we want to estimate the completeness). The positions of these
artificial sources are randomly distributed within the two primary
beams. The input source is considered   recovered when it is
extracted with S/N$\geq 3.5\sigma$. 
Within the selected flux densities range (0.05 to 1 mJy) we iterated the procedure of inserting artificial sources for each continuum map 1000 times, using four to eight artificial sources in each
  simulation for each field.  The completeness calculated in each
map, C$_{i}(S)$, is equal to the ratio between the number of recovered
sources and the number of input sources for each flux $S$.  Figure
\ref{fig:completeness} shows $C_{i}(S)$ as a function of the
intrinsic flux density $S$, estimated on the deepest ALMA continuum map
(Field $a$ in Table~ \ref{tab:field}).

The beam response is not uniform and decreases with increasing  distance from the
map centre. Therefore, the effective area that is sensitive to a given flux $S$ decreases
rapidly with the flux itself. As a consequence, the effective area of the survey
depends on the considered flux, i.e.  $A_{\rm
  survey}(S)$. Since the completeness $C_{i}(S)$ is estimated on the
continuum maps that have been corrected for primary beam attenuation, the
completeness function already automatically includes the effect of
variation of sensitivity as a function of distance from the map
centre. Therefore, the effective survey area of each map at a given
flux $S$ is given by $C_{\rm i}(S)A_{i}(S)$, where $A_{i}(S)$ is the
two-primary-beam area of the ALMA map $i$.  Therefore, the total
effective survey area is given by

$$
A_{\rm survey} (S) = \sum_{i} C_{\rm i}(S)A_{i}(S) ~.
$$ 

\begin{figure}
   \centering
\includegraphics[width =.5\textwidth]{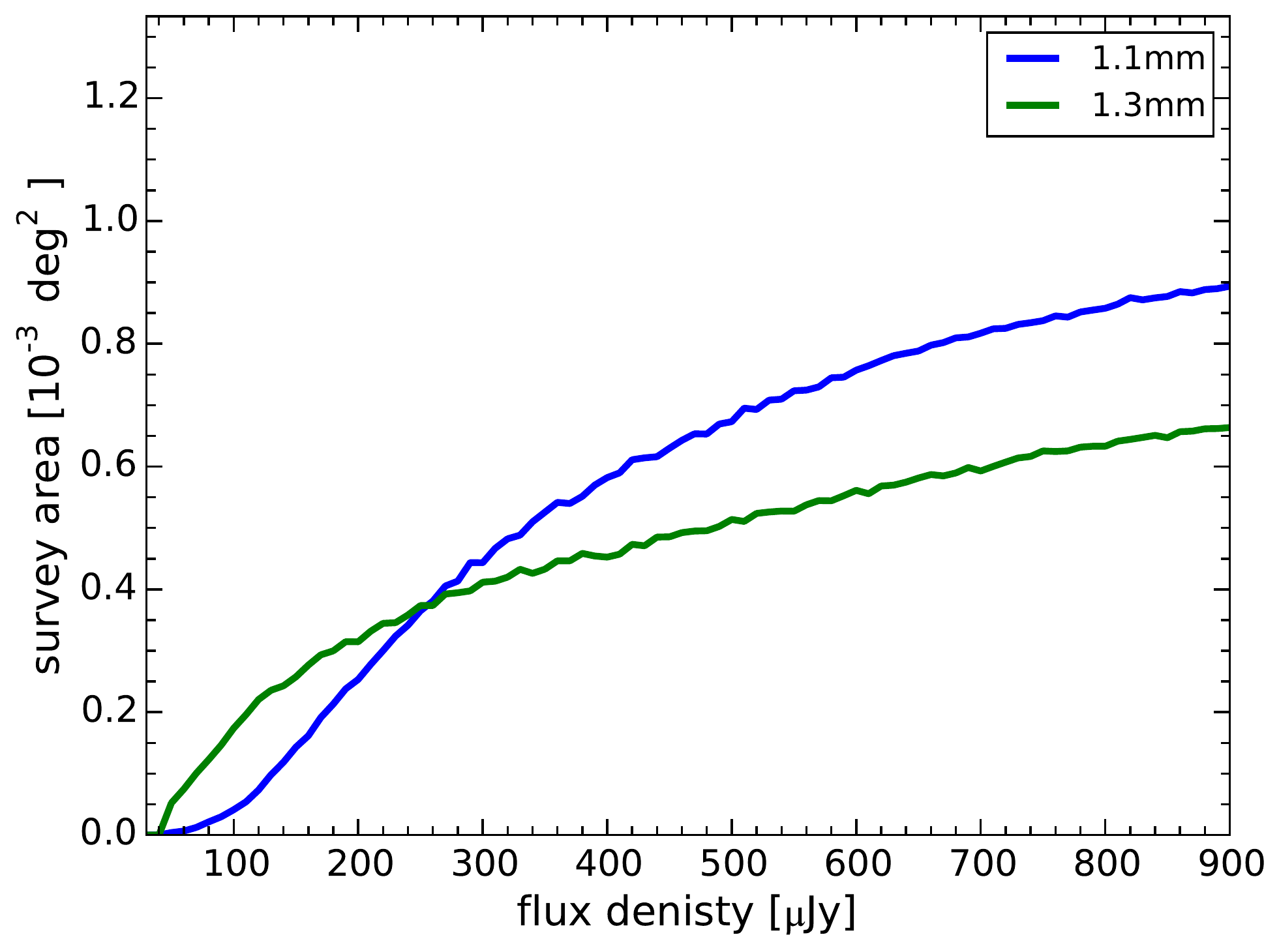}

\caption{Effective survey area as a function of intrinsic flux density. This is the area over
which a source with an intrinsic flux density $S$ can be detected with $S/N >3.5\sigma$. The blue and green
curves
are the survey areas for the maps at 1.1 mm and 1.3 mm, respectively. }
 \label{fig:survey_area}
   \end{figure}

 We obtained  $A_{\rm survey}(S)$ both at 1.1 mm and at 1.3 mm, as shown in Figure \ref{fig:survey_area}.

\subsection{Flux boosting}\label{sec:flux_boosting}
 The noise fluctuations in continuum maps may influence photometric
  measurements of the extracted sources. 
  Since the counts of faint sources increase with decreasing flux density (e.g. \citealt{Scott:2012,Hatsukade:2013,Ono:2014}),  there should be a `sea' of faint source below the noise level that may influence photometric extraction.
  There is, indeed, a greater
  probability that intrinsically faint sources are detected at higher
  flux, rather than that bright ones  are de-boosted to a lower flux. (See \cite{Hogg:1998}  and \cite{Coppin:2005} 
   for a full description of this effect.) This effect,
  called flux boosting, is extremely important at low S/N (< 5)
  where flux measurement can be overestimated.
  Since we do not know the priori distribution of flux densities in the range
  0.001-1 mJy, we performed a simple simulation to estimate the boost factor 
  as a  function of S/N detection.

\begin{figure}
   \centering
\includegraphics[width =.5\textwidth]{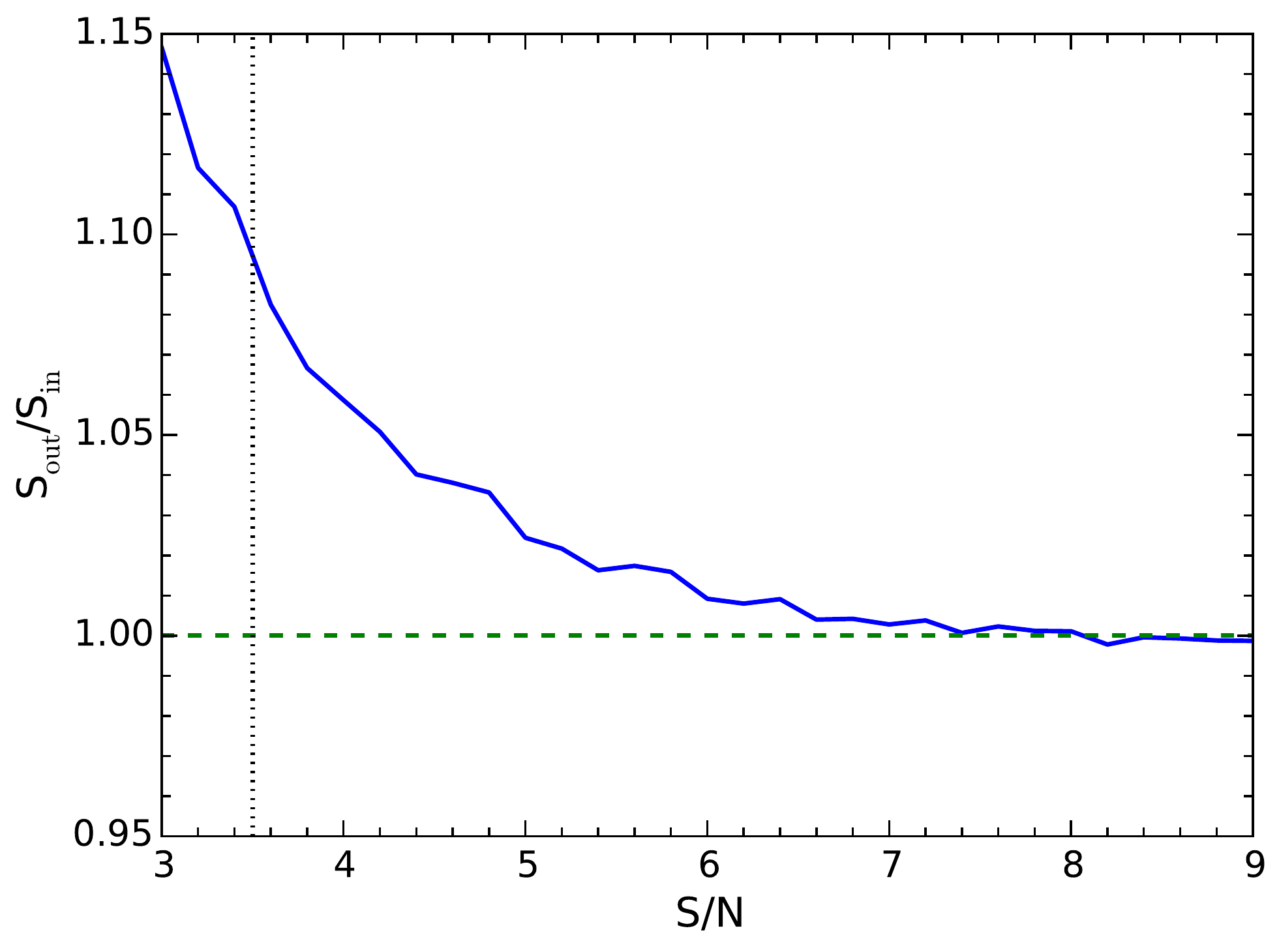}

\caption{Flux-boosting factor as a function of S/N, estimated from simulations.  The horizontal dashed line
corresponds to $S_{\rm out}$ = $S_{\rm in}$ . The vertical dashed line corresponds to the detection threshold, S/N = 3.5. }
 \label{fig:flux_boosting}
   \end{figure}

The simulation was carried out in the map that was uncorrected for primary beam
attenuation. Into the maps we inserted flux-scaled artificial sources
(4-8) whose S/N are in the range of 3-8. Then, we extracted the flux
densities at the same position where the sources were located.  The a priori knowledge of the source position in this
process
may lead to underestimating the flux boosting correction since the noise in the maps may lead
to an offset in the recovered position of mock (or real) sources. However this
effect is maximum when sources are near a noise peak. Since the number of mock
(or real) sources in each continuum map is low (<10), the probability that an
artificial (or real) source is near to a noise peak is lower than 5$\%$. In
conclusion, the  error on the flux-boosting factor, which is due to the a priori knowledge of artificial source position, is  small ($<5\%$).  The flux boosting is calculated as the ratio of the
measured flux density ($S_{\rm out}$) to the input flux density
($S_{\rm in}$). We repeated this simulation for each ALMA continuum map $10^4$ times.  Figure \ref{fig:flux_boosting} shows the average
ratio of the extracted flux densities S$_{\rm out}$ to the input flux
densities S$_{\rm in}$ as a function of S/N. At S/N = 3.5, the boost
fraction is $\simeq$1.09, so the difference between extracted flux and
input flux is less than 10$\%$.  The boosting factor correction (Figure
\ref{fig:flux_boosting}) was then applied to the measured detection
fluxes.

\section{Results}

 To summarise the previous sections,
we detected 50 sources with S/N $>3.5$ in 18 continuum maps in the ALMA
Bands 6 and 7.
Then, we corrected the flux densities for the flux-boosting effect. In the following,
we discuss the additional steps required to infer the sources' number counts and the comparison
with the CIB.

\subsection{Differential number counts}\label{sec:diff}

We scaled the flux density of the sources observed at
$\lambda < 1.2$ mm to the flux density at 1.1 mm and those at $\lambda
> 1.2$ mm to the flux density at 1.3 mm. 
 The reason for splitting the sources into these two wavelength ranges is 
that this significantly reduces the uncertainties on the flux obtained
from the extrapolation. A more detailed analysis of these issues is
given in Appendix \ref{sec:appendix_B}. With this strategy, the error
on the flux associated with each source is always less than 18\%.
Following the prescription of
\cite{Hatsukade:2013} and \cite{Ono:2014}, we estimated the number
counts at two different wavelengths: 1.1 mm and 1.3 mm.
We then estimated the effective survey area associated with the flux of each source.
To estimate the number counts, we corrected for the contamination of
`spurious' sources, i.e. the fraction of sources that are due to noise fluctuations above the
3.5$\sigma$ level (and meeting the additional requirements given in sect 3.1). The contamination
fraction $f_{\rm c}$ was inferred from the fraction of negative sources at
each S/N level, as inferred from Figure \ref{fig:histo}.
For each source,
$f_{\rm c}$ is the ratio between negative and positive detections at
its S/N.
Therefore, the contribution of each source to the number
counts is (1-$f_{\rm c}$) divided by its respective effective survey
area $A_{\rm survey}(S)$.  We carried out the logarithmic differential
number counts $dN(S)/d\Log S$ in logarithmic flux density bins with size $\Delta
\Log S = 0.2$. So the logarithmic differential number counts for a
selected bin is given by

$$
\frac{dN(S)}{d{\rm Log} S}\bigg|_{S\pm 1/2\Delta{\rm Log}S} = \sum_{j} \frac{1-f_{{\rm c}j}}{A_{\rm survey}(S)},
$$

where $j$ are all sources with flux density between Log$S$-1/2$\Delta{\rm
Log}S$ and Log$S$+1/2$\Delta\Log S$. {\rm The resulting differential number counts are scaled to $\Delta{\rm Log}S$ = 1.}

 The total uncertainty on the logarithmic differential number counts is computed by combining the contribution
from Poisson noise, from the cosmic variance and  from errors due to completeness and flux-boosting 
corrections. In the following we evaluate each single contribution:
\begin{itemize}
\item  The observational uncertainty related to the actual number of detected sources is 
calculated from Poisson  confidence limits of 84.13\%
\citep{Gehrels:1986} by using the number of sources detected in each bin.

\item The error due to the cosmic variance is estimated by using a software tool provided by 
\cite{Moster:2011}. This procedure uses predictions of the underlying structure of cold dark 
matter and the expected bias for a galaxy population. The estimate depends on the angular 
dimension of the field, the mean redshift, the redshift bin size, stellar mass of the galaxy 
population in question, and  also on the number of independent fields sampled in different regions of the sky.  We assume a  mean redshift  of $z=3.5$, a redshift 
bin size of  $dz = 3,$ and a stellar mass of 10$^{10.5}$ \msun\ from  \cite{Yun:2012} and
\cite{Weis:2013} who measured and analysed SEDs and redshifts of bright (S$>$1mJy) submillimetre galaxies through 
spectroscopic and photometric observations. Considering that  for widely separated fields the cosmic variance goes as  1/$\sqrt{\rm{N_{\rm field}}}$, the relative error  is $< 18\%$ in the deepest logarithmic differential number count bin.

\item The relative uncertainties relating to completeness and flux-boosting corrections of the order of 5\%.
 \end{itemize}
Because the cosmic variance and errors induced by count estimations (completeness, flux boosting) are less than 
20\%, the uncertainty on logarithmic differential number counts is completely dominated by   the Poisson errors.

\begin{figure*}
        \centering

                \includegraphics[width=0.49\textwidth]{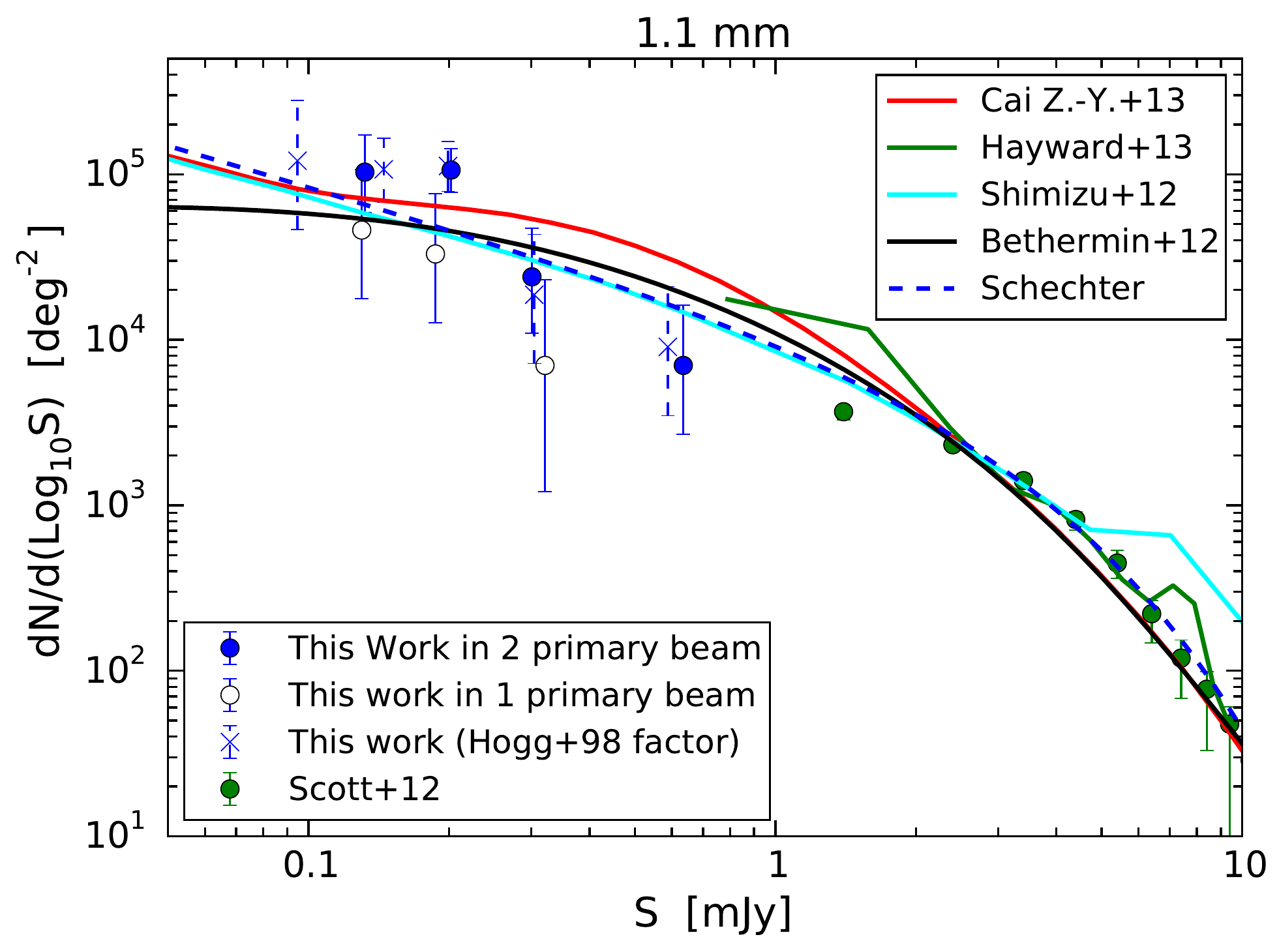}
                \includegraphics[width=0.494\textwidth]{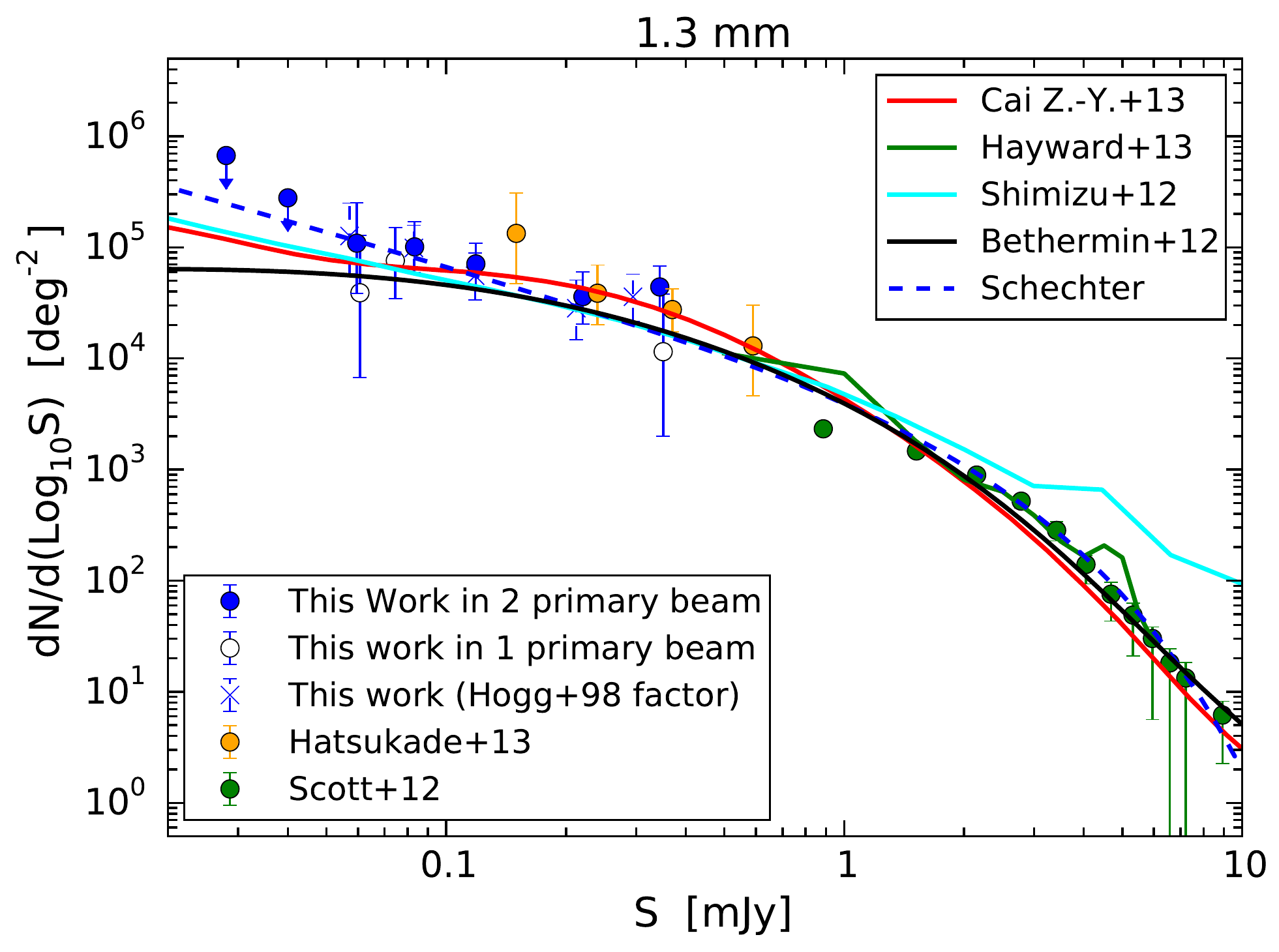}
        \caption{Logarithmic differential number counts as a function of flux density at $\lambda = $1.1 mm and
        $\lambda = $1.3
        mm. The blue solid  and upper limit symbols are the results of this work. The hollow symbols are the results
        obtained by using only the sources within one primary beam. 
        The blue crosses with dashed error bars are the differential number counts corrected for flux boosting
        by using Equation 4 in \cite{Hogg:1998}. The green symbols are estimated from
        \cite{Scott:2012}. The orange symbols are the number counts estimated by \cite{Hatsukade:2013}.
         The red, green, cyan, and black solid curves are the model predictions obtained by \cite{Cai:2013}, \cite{Hayward:2013}, \cite{Shimizu:2012}, and \cite{Bethermin:2012}, respectively. The blue dashed curve shows the best-fit Schechter function.}
 \label{fig:log_dn_ds}
   \end{figure*}

The resulting differential number counts are summarised in Table
\ref{tab:diff} and shown in Figure \ref{fig:log_dn_ds} (blue solid symbols). Number counts could be
derived down to 60$\mu$Jy at 1.3mm and down to 100$\mu$Jy at 1.1mm. Moreover, since we 
do not detect any faint sources with flux densities $30\lesssim S \lesssim 50$ $\mu$Jy in the deepest ALMA map
(Field \emph{a} in Table \ref{tab:field}) with sensitivity of $\sim7.8$ $\mu$Jy,  we can set a tight upper limit on the
number counts at S~=~30~$\mu$Jy and  at S~=~40~$\mu$Jy.  We note that, with the latter, we constrain the number counts
at flux levels that are a factor of four deeper than previous studies \citep{Ono:2014}. We also show separately
(hollow symbols) the number counts inferred by only using the sources
 detected within the primary beam (i.e. 7
  sources at 1.1 mm and 6 sources at 1.3 mm). In the latter case, the statistical errorbars
  are obviously larger but fully consistent (within errors)
  with the number counts inferred over two beams.

\begin{table}
\caption{Differential number counts.}           
\label{tab:diff}      
\centering          
\begin{tabular}{l c c }    
 & $\lambda = $ 1.1 mm & \\
\hline
\hline
 S [mJy]                &       $dN/d\Log(S)$ [10$^4$]           &      N$_{\rm detections}$    \\
\hline

\\
0.13    &       $10_{-4}^{+7}$          &       5               \\
0.20    &       $11_{-3}^{+3}$          &       14              \\
0.30    &       $2_{-1}^{+2}$           &       3               \\
0.63    &       $0.7_{-0.4}^{+0.9}$ &   2               \\ 
\hline   

\\
&  $\lambda = $ 1.3 mm  &\\
\hline
\hline
 S [mJy]                &       d$N$/d$\Log(S)$ [10$^4$]         &      N$_{\rm detections}$    \\
\hline

\\
0.03 &    <70 & - \\
0.04 &    <30 & - \\
0.06    &       $11^{+14}_{-7}$         &       2               \\
0.08    &       $10^{+7}_{-4}$          &       5               \\
0.12    &       $7^{+4}_{-3}$           &       7               \\
0.22    &       $3^{+2}_{-2}$           &       5               \\
0.34    &       $4^{+2}_{-2}$           &       7               \\ 
\hline   

\end{tabular}   
\end{table}

%
%

We also show the number counts of bright (S > $1$ mJy) SMGs obtained by \citet{Scott:2012} at 1.1~mm
with AzTEC. However, the two faintest bins in the latter data are not
considered when comparing models or when fitting analytic functions since
the completeness at these flux densities is too low and the number
counts are underestimated.
Since there are no number counts of bright sources at 1.3 mm, we
 used the number counts at 1.1 mm by scaling the flux density to 1.3
 mm flux density.  Figure \ref{fig:log_dn_ds} shows that the
 differential number counts increase with decreasing flux density. At
 1.3 mm, the differential number counts of \citet{Hatsukade:2013} (orange symbols), which are obtained from sources with S/N$\geq4$, are
 consistent with our results within the uncertainties, but our slope
 of the logarithmic number counts at sub-mJy flux densities is flatter
 than those obtained by \citet{Hatsukade:2013}.   We do not plot
   the number counts of \cite{Ono:2014} as they were estimated at 1.2
   mm using continuum maps over the whole 1.04-1.22 mm wavelength range, i.e. they were not
   extracted with the same procedure we used to define our sample.  Given that there is no information
   yet on the SED or on the redshift of the sources contributing to the number counts, 
   our approach of dividing sources into two wavelength ranges, hence minimising the
   extrapolation assumptions, provides more solid results, as discussed  in   more detail
   in Appendix \ref{sec:appendix_B}.

 Since some previous works assess the boosting factor correction by using a Bayesian estimation (e.g. \citealt{Coppin:2005}) instead of that shown in Sec. \ref{sec:flux_boosting}, we verified that the final results do not depend on the method used to correct the flux-boosting effect. 
Therefore, we  show (with crosses and dashed error bars)
the differential number counts by correcting for
a boost factor that is estimated from Equation (4) of \cite{Hogg:1998} (note that the
correction was on the data {\it not} corrected for flux boosting as described
in Sect. \ref{sec:flux_boosting}, otherwise this would result in a double
correction). To apply the prescription of \cite{Hogg:1998},
we used the best-fit Schechter function (Sect. \ref{sec:schechter_function})
as a conservative
a priori distribution of flux densities at faint fluxes (we shall see
that a
Schechter function tends to give an  extrapolation of the number counts
that is steeper than expected for
models and tends to overproduce the CIB).
It should, however, be noted that
the prescription given by \cite{Hogg:1998} may not apply to these data,
since the noise is not uniformly distributed over the field of view as
a consequence of the primary beam attenuation.
However, as seen in Figure \ref{fig:log_dn_ds}, the slopes obtained with this correction factor
are consistent with those obtained  
using a boost factor estimated in Sec. \ref{sec:flux_boosting}.

We also note that, while this paper was under review, a paper was posted on arXiv in which differential number
counts down to 20$\mu$Jy are estimated by exploiting data on a lensing cluster \citep{Fujimoto:2015}. Their claimed
number counts at 20-40$\mu$Jy are significantly higher then the upper limits estimated by us on unlensed sources.
We tentatively ascribe the discrepancy to uncertainties associated with the calculation of the lensing
factor for sources of unknown redshift and to uncertainties associated with the complex calculation of the
effective survey area in the presence of strong lensing. However, new forthcoming deep ALMA observations will
enable us to  clarify these discrepancies further.

\subsection{Comparison with models}
We compared the differential and integrated number counts to the
theoretical results obtained by recent simulations and semianalytical
models.  In Figure \ref{fig:log_dn_ds}, the differential number counts
by \cite{Hayward:2013} are shown with a green line.  Their results
were obtained by combining a semi-analytical model with 3-D
hydrodynamical simulations and 3-D dust radiative transfer
calculations. The main contributions to the mm counts is from
isolated-disc, galaxy pairs, and late-stage merger-induced
starbursts. Figure \ref{fig:log_dn_ds} shows that their model
predictions are able to reproduce the observational results at
flux densities $ > 2$ mJy,


We also compared our results with the model by
\citet{Shimizu:2012}, who performed cosmological hydrodynamics
simulations using an updated version of the Tree-PM smoothed particle
hydrodynamics code, GADGET-3. They assume feedback mechanisms
were triggered by supernovae and the SED of galaxies at mm-FIR
wavelengths were described by a modified black body emission. This
model predicts that the bright SMGs reside in greater massive halos
($>10^{12}$ \msun) and that their typical stellar masses are greater
than $10^{11}$ \msun. Their results are broadly consistent with the
ALMA results up to 1--5 mJy. However, their estimated number counts of
bright SMGs ($>1-5$ mJy) are significantly higher than the observed
number counts, both at 1.1 and 1.3 mm. According to this model,  
approximately 90\%  of millimeter sources in the flux range of 0.1-1 mJy
are at $z>2$. Therefore, most of the observed sources are high-$z$
galaxies and the contribution from low-$z$ is small. 

 \cite{Bethermin:2012} developed an empirical model in which they start
from mid-IR and radio number counts, and by using a library of SEDs, they
predict the
number counts at far-IR and millimeter wavelengths.
This model is based on a 
redshift evolution of the SEDs associated with the
two star formation modes: main-sequence and starburst. 
The predictions of their empirical model are plotted 
as solid black curves in both panels of Figure~\ref{fig:log_dn_ds}. Their corresponding SEDs are derived from Herschel observations. 
The predictions are slightly below  the observed faint-end, both at 1.1 mm and 1.3 mm. 
However, the generally good matching of the model with the observations
suggests that the faint 
millimeter sources (S $<$ 1mJy) are more likely associated with normal (main
sequence) star-forming galaxies, since
the starburst emission dominates at higher flux densities at these wavelengths. 

Finally, Figure \ref{fig:log_dn_ds} shows the theoretical predictions
of \cite{Cai:2013}\footnote{The models predictions are available in electronic
format at the Web site http://staff.ustc.edu.cn/$\sim$zcai/}. 
The \cite{Cai:2013} model starts by considering the
observed dichotomy in the ages of stellar populations of massive
spheroidal (components of) galaxies on one side and late-type galaxies
on the other. Spheroidal galaxies and massive bulges of Sa-type galaxies must have
formed most of their stars at $z\gtrsim 1$, while the disc components
of spirals and the irregular galaxies are characterised by
significantly younger stellar populations, with star-formation activity
continuing up to the present time.  
The model
includes a self-consistent treatment of the chemical evolution of the
ISM, calculated using the standard equations and stellar
nucleosynthesis prescriptions. The chemical evolution controls the
evolution of the dust abundance, hence the dust absorption and
re-emission.  On the other hand, the evolution of late-type galaxies
was described using a phenomenological approach and considering two
populations with different SEDs and
different evolutionary properties: `normal' late-type galaxies, with
low evolution and low dust temperatures (`cold' population) and
rapidly evolving starburst galaxies, with warmer dust temperatures
(`warm' population).  Their results are in good agreement with our
differential number counts from faint to bright flux densities at 1.3
mm and also 1.1 mm (although with some deviations). 
According to this model the steep slope of the bright counts is accounted 
for by the sudden  appearance of star-forming proto-spheriodal galaxies 
at $z\gtrsim 1.5$, whose counts already begin to flatten at flux densities 
of a few-to-several mJy's. The counts of starburst 
galaxies have a somewhat flatter slope and come up at levels similar
to those of proto-spheroids at the flux densities of the (new) faint 
counts, while the contribution of `normal' late-type galaxies is 
always minor in the considered flux density range  but increases with
decreasing flux density. The redshift distribution at the flux densities 
of the present counts is bimodal, with starburst galaxies peaking at 
$z\simeq 1.5$ and proto-spheroids peaking at $z\simeq 2$. At bright 
flux densities ($S\sim 10\,$mJy), the starburst galaxy peak shifts to $z
\ll 1$ (being the brightest, in flux terms, starburst galaxies are mostly 
local) while the proto-spheroid one remains at $z\simeq 2$.




\subsection{Source counts parametrisation with a Schechter function}\label{sec:schechter_function}

We also parametrised the differential number counts using a Schechter
function:

$$
\frac{dN}{dS} dS = \phi_{\star}\left(\frac{S}{S_\star}\right)^\alpha\exp\left(-\frac{S}{S_\star}\right)d\left(\frac{S}{S_\star}\right)
,$$ with $\phi_\star$ being the normalisation, $S_\star$ the
characteristic flux density, and $\alpha$  the faint-end slope of the
number counts. We fitted  the Schechter function separately at 1.1 mm
and 1.3 mm by using the number counts estimated in this work and from
the literature (e.g. \citealt{Scott:2012},
\citealt{Hatsukade:2013}). We did not use the two faintest
data points from \cite{Scott:2012} because they may suffer from
completeness problems. 
The three free parameters
were then derived by $\chi^2$ minimisation. Table \ref{tab:schechter}
reports the best-fit parameters and Figures \ref{fig:log_dn_ds} and
\ref{fig:integrated} show the results of the Schechter function
fitting.

The reduced $\chi^2$ are 0.9 at 1.1 mm and 1.1 at 1.3 mm,
 meaning that the differential number counts can be properly
described by a Schechter function down to the flux levels observed by us.
The two faint-end slopes are
similar within the errors, suggesting that the two number counts can be
described by the same function. The bright-end shape also
  matches with a pure Schechter function perfectly well at both wavelengths, which has recently been observed in \cite{Dayal:2014}. 

However, we note that the slope of the Schechter function is significantly
steeper than expected by models (especially at 1.3 mm) and would overproduce
the CIB at faint fluxes, even taking the upper boundary of the CIB, which
is discussed in the next section. Therefore, we warn that the
Schechter function fitted to the current data is probably not suitable
for describing the number counts at fluxes fainter than those observed by us.



\begin{table}
\caption{Best-fit parameters of the Schechter function at $\lambda =$1.1,, and $\lambda =$1.3mm.}           
\label{tab:schechter}      
\centering          
\begin{tabular}{c c c c }    
\hline
\hline
 $\lambda$ & $\phi_\star$ [deg$^{-2}$]  &       $S_\star$ [mJy]  &      $\alpha$        \\
\hline

\\
1.1 mm  &       $(2.7\pm0.9)\times10^{3}$ &  $2.6\pm0.4$  & $-1.81\pm0.14$              \\
1.3 mm  &       $(1.8\pm0.4)\times10^{3}$ &  $1.7\pm0.2$  & $-2.08\pm0.11$              \\
\hline   

\end{tabular}   
\end{table}

\subsection{Cosmic infrared background}

\begin{figure*}
        \centering

                \includegraphics[width=0.49\textwidth]{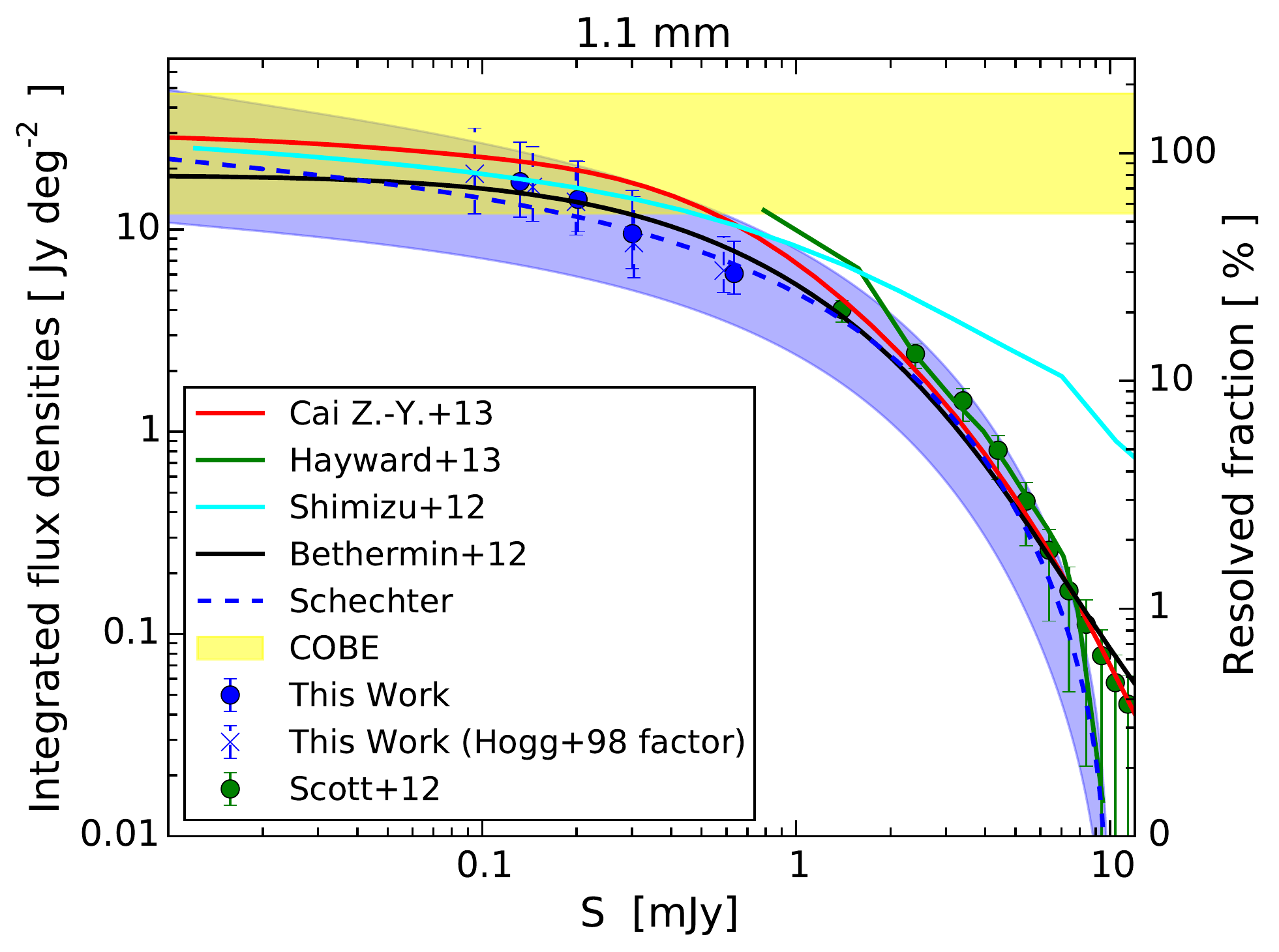}
                \includegraphics[width=0.49\textwidth]{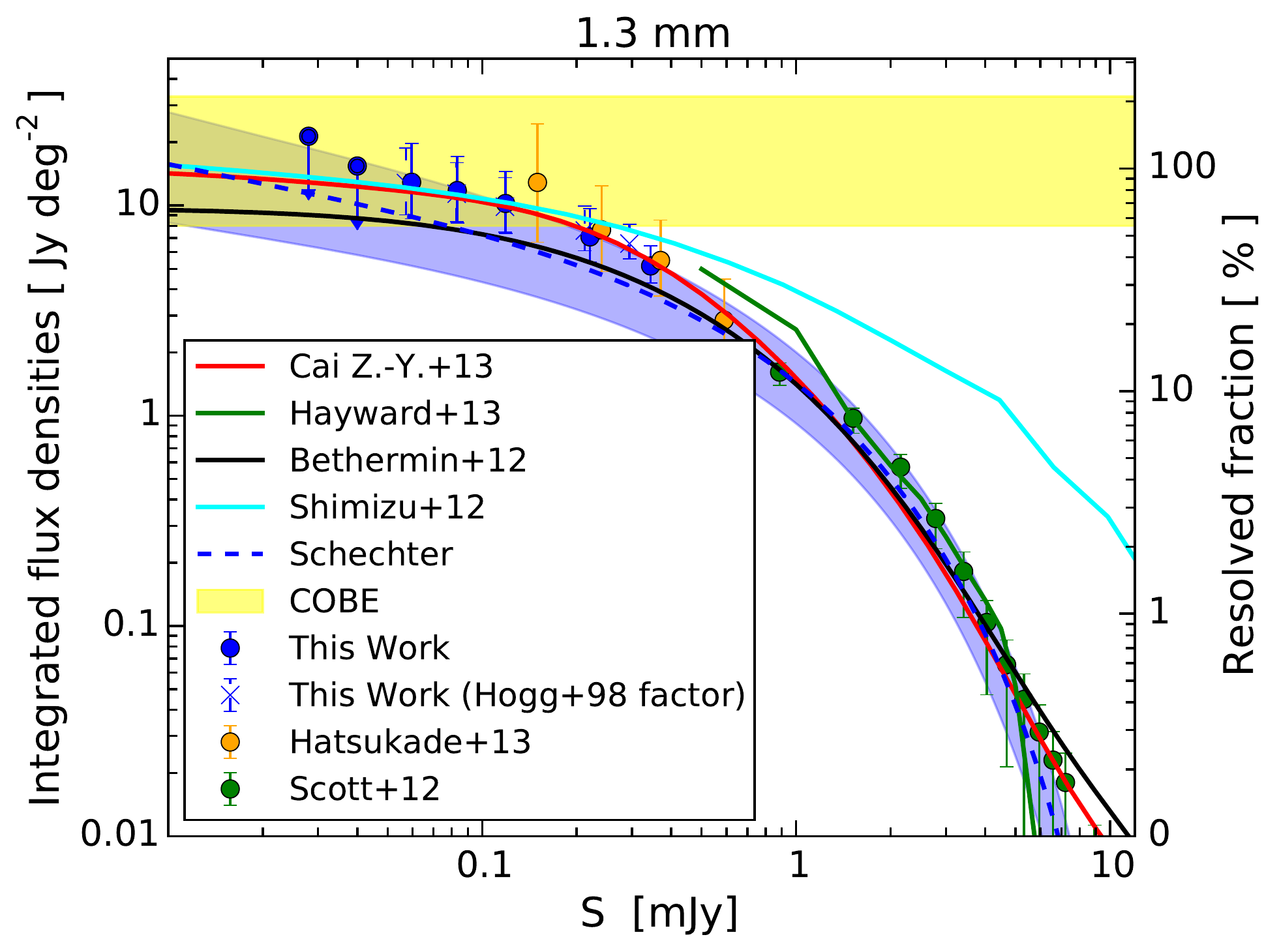}

        \caption{Integrated flux density at $\lambda = $1.1 mm and $\lambda = $1.3 mm. The
          right axis shows the fraction of resolved CIB. The yellow
          shaded region is the CIB measured by COBE
          \citep{Puget:1996,Fixsen:1998}. The blue symbols are the
          results from this work  (we used the differential number counts by   
          \cite{Scott:2012}, indicated by green symbols for $S >$ 1mJy ). The blue crosses with dashed error bars are the
           results from this work, corrected for the flux boosting, using
           Equation 4 in \cite{Hogg:1998}.
          The orange symbols are from the number counts
          estimated by \cite{Hatsukade:2013}.  The red, green, cyan, and black solid curves are the model predictions  by \cite{Cai:2013}, \cite{Hayward:2013}, \cite{Shimizu:2012}, and \cite{Bethermin:2012}, respectively.  The blue dashed curve
          shows the integrated flux densities estimated from the best-fit Schechter function, and
          the blue shaded region indicates its statistical uncertainty.} 
 \label{fig:integrated}
   \end{figure*}

We calculated the 1.1 mm and 1.3 number counts down to 60 $\mu$Jy
using Cycle 0 and 1 ALMA observations. 
Using the improved number counts estimated in this work, 
we estimated the integrated flux densities
from resolved sources and we derived the fraction of the CIB resolved
by ALMA at 1.1 mm and 1.3 mm.  The integrated flux densities are given
by

$$
I(S>S_{lim}) = \int^{\infty}_{S_{\rm limit}} \frac{dN(S)}{dS} S dS, 
$$

where 

$$
\frac{dN(S)}{dS} =  \frac{dN(S)}{dLogS}\frac{1}{S \ln(10).}
$$

Figure \ref{fig:integrated} shows the integrated flux densities at 1.1
mm and 1.3 mm as a function of $S_{\rm limit}$.  We note that we used the results
of \cite{Scott:2012} at bright flux ($S >$ 1mJy), but excluding the two point
at faintest fluxes, because of incompleteness issues. The integrated counts
down to $S_{\rm limit} = 0.1 $ mJy at 1.1 mm and to $S_{\rm limit} =
0.06$ mJy at 1.3 mm are $17^{+10}_{-5}$ Jy deg$^{-2}$ and
$13^{+6}_{-3}$ Jy deg$^{-2}$ respectively.  
We compared these results with the analytical fit obtained by
\cite{Fixsen:1998} from the COBE measurement: $25_{-13}^{+22}$ Jy deg$^{-2}$
at 1.1 mm and $17_{-9}^{+16}$ Jy deg$^{-2}$  at 1.3 mm (see also \citealt{Lagache:1999}
and \citealt{Schmidt:2015}). 
Since these
measurements suffer from large uncertainties (especially due to the
uncertainties on  the Galactic contribution), we are not able to exactly determine
 the fraction of CIB resolved. Certainly, even by taking the highest
value of the CIB  that is consistent with the uncertainties given by \cite{Fixsen:1998},
we can say that at 60$\mu$Jy  more than 50\% (but probably much more) of
the CIB is resolved.
Thus  our results provide a   lower limit on the CIB intensity at 1.1 mm and 1.3 mm. Moreover, the flatness of the faint end-slopes, in particular the flatness at 1.3 mm, suggests that the integrated flux densities estimated in this work are likely to be close to the CIB intensity.

The blue curve gives the integrated number counts inferred from the Schechter
function that fit the differential number counts, and the shaded blue area gives
the associated uncertainty. The uncertainty of the latter is large enough to be
consistent with any value of the CIB within the range given by
\cite{Fixsen:1998}, however the slope would indicate that this functional form
would saturate even the highest boundary at the CIB at faint fluxes, as
discussed above.

We note
that the integrated number counts show a clear flattening at lowest flux bins populated by
our detections. The flattening of the number counts is  supported 
by the tight upper limits at the faintest fluxes sampled by us. Such
flattening of the number counts, fully consistent with the models,
suggests that we are actually resolving most of the CIB at our faint fluxes.


It will be of great interest to investigate, with followup observations, what 
the redshift distribution is of these sources that produce most of the background,
especially at the faint end. In the meantime, we can infer what the properties are of these galaxies in terms of SFR. Indeed, because of the negative K-correction,
at 1.3mm a given observed flux
density corresponds to an IR-luminosity (hence a SFR) that is nearly independent
of redshift, across the entire redshift range 0.5$<z<$15 \citep[by adopting
the conversion factor and IMF given in ][]{Blain:1999, Maiolino:2008}.
In particular, 
the minimum flux density sampled by us, 60 $\mu$Jy at 1.3mm, corresponds to a total IR
luminosity $\rm L(8-1000\mu m)=6~\times10^{44}~erg~s^{-1}$, corresponding to a $\rm
SFR=40~M_{\odot}~yr^{-1}$ \citep{Kennicutt:2012}, nearly independent of redshift.
 This is consistent with the predictions of \cite{Bethermin:2012} and \cite{Cai:2013} as 
they expect  faint sources to be associated with `normal' star-forming galaxies.
We can therefore state that galaxies with $\rm SFR<40~M_{\odot}~yr^{-1}$ certainly
contribute less than 50\% of the CIB at 1.3 mm, and probably a much lower percentage. 
Vice versa, the bulk of
the CIB (50\% and probably much more) must be due to galaxies
with $\rm SFR>40~M_{\odot}~yr^{-1}$.

\section{Summary and conclusions}

We have used 18 deep ALMA maps in Bands 6 and 7 to investigate the number counts of sources
at millimeter wavelengths. The sensitivity (rms)
of these ALMA maps range from 7.8~ $\mu$Jy/beam to 52 $\mu$Jy/beam.

Sources were detected down to 3.5$\sigma$. As a requirement
for detection, we applied that the  size of the sources should be equal to the beam size
(or slightly larger) within the uncertainties. Since the noise due to bad
UV-visibilities, or to
sidelobes emissions from bright sources, or to thermal noise associated with individual
antennae or group of antennae 
should introduce fluctuations that have a spatial shape that is completely different from the coherent
beam, this criterion enables us to remove most of the spurious detections associated with
noise fluctuations.

We searched for sources out to a distance equal to the diameter of two primary beams. However, we have checked
that the final number counts do not change, within errors, by restricting the source search
to within the primary beam (although the statistics are obviously lower).

A total of 50 sources were detected that match these criteria.
We explored counts at these two different wavelengths, 1.1mm and 1.3mm (hence the ALMA maps were
divided into two  groups, depending on their central wavelength). This approach
avoids large flux extrapolation from observations obtained at different wavelengths. Since
we do not yet know  the intrinsic SED and redshift distribution of the detected sources (which
would be required for a proper extrapolation of the fluxes from different wavelengths), our
approach provides safer results, although at the expense of lower statistics.

Number counts were obtained by taking into account completeness, flux-boosting effects,
correction for spurious sources, and an effective survey area at different flux limits.
We extracted differential number counts down to 60 $\mu$Jy and 100 $\mu$Jy at 1.3mm and 1.1mm,
respectively, inferring sources' number densities of $\rm dN/d(\Log S)\sim 10^5 ~deg^{-2}$
at these faint limits. Using the deepest ALMA field, we inferred tight upper limits on the number 
counts at 30 $\mu$Jy and at 40 $\mu$Jy.

The differential number counts at 1.1mm (1.3mm), across the entire range from 60 $\mu$Jy to 10 mJy,
can be fitted with a Schechter function with a faint
end slope $\rm \alpha \approx -1.8$ ($\rm \alpha \approx -2.0$),
a characteristic flux density $\rm S_*\approx 2.6~ mJy$ ($\rm S_*\approx 1.7~mJy$), and
a normalisation at $\rm S_*$ of $\rm \phi_*=2.7\times 10^3~deg^{-2}$
($\rm \phi_*=1.8\times 10^3~deg^{-2}$). 
 We note that these Schechter
functions describe the number counts down to 60--100 $\mu$Jy, but their extrapolation
to fainter fluxes is not trustworthy. 

 The large uncertainties affecting our knowledge of the CIB level prevents us from setting 
tight limits on the fraction that is resolved by our data. Clearly, at least 50\% of the
CIB is resolved by our data (and probably much more). However, our results
set a  lower limit to the
 CIB intensity at 1.1-1.3 mm, significantly above the one coming from 
 direct measurements. The flatness of the faint counts implies that 
 this lower limit is likely to be close to the CIB intensity.
%

The SFR of the sources contributing to the CIB at such faint fluxes is about
$\rm 40 ~M_{\odot}~yr^{-1}$, independent of their redshift. We therefore infer that
sources with $\rm SFR < 40 ~M_{\odot}~yr^{-1}$ contribute less than half of the CIB at
1.3~mm, and probably much less. Conversely, the bulk of the CIB must be produced by
galaxies with $\rm SFR > 40 ~M_{\odot}~yr^{-1}$.

\begin{acknowledgements}
We thank the anonymous referee for comments and suggestions that improved the paper. We thank Paola Andreani, Matthew Auger and Valentina Calvi for comments and discussions.   This paper makes use of the 
following ALMA data: ADS/JAO.ALMA\#2011.1.00115.S, ADS/JAO.ALMA\#2011.1.00243.S, 
ADS/JAO.ALMA\#2011.1.00767.S, ADS/JAO.ALMA\#2012.1.00142.S, ADS/JAO.ALMA\#2012.1.00374.S, 
ADS/JAO.ALMA\#2012.1.00523.S, ADS/JAO.ALMA\#2012.1.00719.S and ADS/JAO.ALMA\#2012.A.00040.S.
ALMA is a partnership of ESO (representing its member 
states), NSF (USA) and NINS (Japan), together with NRC (Canada) and NSC and ASIAA (Taiwan), in 
cooperation with the Republic of Chile. The Joint ALMA Observatory is operated by ESO, AUI/NRAO and
 NAOJ. GDZ and MN acknowledge financial support from ASI/INAF Agreement 2014-024-R.0 for the {\it Planck} LFI activity of Phase E2. \end{acknowledgements}

\bibliographystyle{aa} 
\bibliography{bibliography_serendipity}

\begin{thebibliography}{47}
\expandafter\ifx\csname natexlab\endcsname\relax\def\natexlab#1{#1}\fi

\bibitem[{{B{\'e}thermin} {et~al.}(2012){B{\'e}thermin}, {Daddi}, {Magdis},
  {Sargent}, {Hezaveh}, {Elbaz}, {Le Borgne}, {Mullaney}, {Pannella}, {Buat},
  {Charmandaris}, {Lagache}, \& {Scott}}]{Bethermin:2012}
{B{\'e}thermin}, M., {Daddi}, E., {Magdis}, G., {et~al.} 2012, \apjl, 757, L23

\bibitem[{{Blain} {et~al.}(1999){Blain}, {Moller}, \& {Maller}}]{Blain:1999}
{Blain}, A.~W., {Moller}, O., \& {Maller}, A.~H. 1999, \mnras, 303, 423

\bibitem[{{Blain} {et~al.}(2002){Blain}, {Smail}, {Ivison}, {Kneib}, \&
  {Frayer}}]{Blain:2002}
{Blain}, A.~W., {Smail}, I., {Ivison}, R.~J., {Kneib}, J.-P., \& {Frayer},
  D.~T. 2002, \physrep, 369, 111

\bibitem[{{Cai} {et~al.}(2013){Cai}, {Lapi}, {Xia}, {De Zotti}, {Negrello},
  {Gruppioni}, {Rigby}, {Castex}, {Delabrouille}, \& {Danese}}]{Cai:2013}
{Cai}, Z.-Y., {Lapi}, A., {Xia}, J.-Q., {et~al.} 2013, \apj, 768, 21

\bibitem[{{Casey} {et~al.}(2014){Casey}, {Narayanan}, \&
  {Cooray}}]{Casey:2014a}
{Casey}, C.~M., {Narayanan}, D., \& {Cooray}, A. 2014, \physrep, 541, 45

\bibitem[{{Chen} {et~al.}(2013{\natexlab{a}}){Chen}, {Cowie}, {Barger},
  {Casey}, {Lee}, {Sanders}, {Wang}, \& {Williams}}]{Chen:2013a}
{Chen}, C.-C., {Cowie}, L.~L., {Barger}, A.~J., {et~al.} 2013{\natexlab{a}},
  \apj, 762, 81

\bibitem[{{Chen} {et~al.}(2013{\natexlab{b}}){Chen}, {Cowie}, {Barger},
  {Casey}, {Lee}, {Sanders}, {Wang}, \& {Williams}}]{Chen:2013}
{Chen}, C.-C., {Cowie}, L.~L., {Barger}, A.~J., {et~al.} 2013{\natexlab{b}},
  \apj, 776, 131

\bibitem[{{Coppin} {et~al.}(2006){Coppin}, {Chapin}, {Mortier}, {Scott},
  {Borys}, {Dunlop}, {Halpern}, {Hughes}, {Pope}, {Scott}, {Serjeant}, {Wagg},
  {Alexander}, {Almaini}, {Aretxaga}, {Babbedge}, {Best}, {Blain}, {Chapman},
  {Clements}, {Crawford}, {Dunne}, {Eales}, {Edge}, {Farrah}, {Gazta{\~n}aga},
  {Gear}, {Granato}, {Greve}, {Fox}, {Ivison}, {Jarvis}, {Jenness}, {Lacey},
  {Lepage}, {Mann}, {Marsden}, {Martinez-Sansigre}, {Oliver}, {Page},
  {Peacock}, {Pearson}, {Percival}, {Priddey}, {Rawlings}, {Rowan-Robinson},
  {Savage}, {Seigar}, {Sekiguchi}, {Silva}, {Simpson}, {Smail}, {Stevens},
  {Takagi}, {Vaccari}, {van Kampen}, \& {Willott}}]{Coppin:2006}
{Coppin}, K., {Chapin}, E.~L., {Mortier}, A.~M.~J., {et~al.} 2006, \mnras, 372,
  1621

\bibitem[{{Coppin} {et~al.}(2005){Coppin}, {Halpern}, {Scott}, {Borys}, \&
  {Chapman}}]{Coppin:2005}
{Coppin}, K., {Halpern}, M., {Scott}, D., {Borys}, C., \& {Chapman}, S. 2005,
  \mnras, 357, 1022

\bibitem[{{Cowie} {et~al.}(2002){Cowie}, {Barger}, \& {Kneib}}]{Cowie:2002}
{Cowie}, L.~L., {Barger}, A.~J., \& {Kneib}, J.-P. 2002, \aj, 123, 2197

\bibitem[{{Dayal} {et~al.}(2014){Dayal}, {Ferrara}, {Dunlop}, \&
  {Pacucci}}]{Dayal:2014}
{Dayal}, P., {Ferrara}, A., {Dunlop}, J.~S., \& {Pacucci}, F. 2014, \mnras,
  445, 2545

\bibitem[{{Dole} {et~al.}(2006){Dole}, {Lagache}, {Puget}, {Caputi},
  {Fern{\'a}ndez-Conde}, {Le Floc'h}, {Papovich}, {P{\'e}rez-Gonz{\'a}lez},
  {Rieke}, \& {Blaylock}}]{Dole:2006}
{Dole}, H., {Lagache}, G., {Puget}, J.-L., {et~al.} 2006, \aap, 451, 417

\bibitem[{{Eales} {et~al.}(1999){Eales}, {Lilly}, {Gear}, {Dunne}, {Bond},
  {Hammer}, {Le F{\`e}vre}, \& {Crampton}}]{Eales:1999}
{Eales}, S., {Lilly}, S., {Gear}, W., {et~al.} 1999, \apj, 515, 518

\bibitem[{{Fixsen} {et~al.}(1998){Fixsen}, {Dwek}, {Mather}, {Bennett}, \&
  {Shafer}}]{Fixsen:1998}
{Fixsen}, D.~J., {Dwek}, E., {Mather}, J.~C., {Bennett}, C.~L., \& {Shafer},
  R.~A. 1998, \apj, 508, 123

\bibitem[{{Fujimoto} {et~al.}(2015){Fujimoto}, {Ouchi}, {Ono}, {Shibuya},
  {Ishigaki}, \& {Momose}}]{Fujimoto:2015}
{Fujimoto}, S., {Ouchi}, M., {Ono}, Y., {et~al.} 2015, ArXiv e-prints
  [\eprint[arXiv]{1505.03523}]

\bibitem[{{Gehrels}(1986)}]{Gehrels:1986}
{Gehrels}, N. 1986, \apj, 303, 336

\bibitem[{{Greve} {et~al.}(2012){Greve}, {Vieira}, {Wei{\ss}}, {Aguirre},
  {Aird}, {Ashby}, {Benson}, {Bleem}, {Bradford}, {Brodwin}, {Carlstrom},
  {Chang}, {Chapman}, {Crawford}, {de Breuck}, {de Haan}, {Dobbs}, {Downes},
  {Fassnacht}, {Fazio}, {George}, {Gladders}, {Gonzalez}, {Halverson},
  {Hezaveh}, {High}, {Holder}, {Holzapfel}, {Hoover}, {Hrubes}, {Johnson},
  {Keisler}, {Knox}, {Lee}, {Leitch}, {Lueker}, {Luong-Van}, {Malkan},
  {Marrone}, {McIntyre}, {McMahon}, {Mehl}, {Menten}, {Meyer}, {Montroy},
  {Murphy}, {Natoli}, {Padin}, {Plagge}, {Pryke}, {Reichardt}, {Rest},
  {Rosenman}, {Ruel}, {Ruhl}, {Schaffer}, {Sharon}, {Shaw}, {Shirokoff},
  {Stalder}, {Stanford}, {Staniszewski}, {Stark}, {Story}, {Vanderlinde},
  {Walsh}, {Welikala}, \& {Williamson}}]{Greve:2012}
{Greve}, T.~R., {Vieira}, J.~D., {Wei{\ss}}, A., {et~al.} 2012, \apj, 756, 101

\bibitem[{{Hatsukade} {et~al.}(2013){Hatsukade}, {Ohta}, {Seko}, {Yabe}, \&
  {Akiyama}}]{Hatsukade:2013}
{Hatsukade}, B., {Ohta}, K., {Seko}, A., {Yabe}, K., \& {Akiyama}, M. 2013,
  \apjl, 769, L27

\bibitem[{{Hayward} {et~al.}(2013){Hayward}, {Narayanan}, {Kere{\v s}},
  {Jonsson}, {Hopkins}, {Cox}, \& {Hernquist}}]{Hayward:2013}
{Hayward}, C.~C., {Narayanan}, D., {Kere{\v s}}, D., {et~al.} 2013, \mnras,
  428, 2529

\bibitem[{{Hogg} \& {Turner}(1998)}]{Hogg:1998}
{Hogg}, D.~W. \& {Turner}, E.~L. 1998, \pasp, 110, 727

\bibitem[{{Johansson} {et~al.}(2011){Johansson}, {Sigurdarson}, \&
  {Horellou}}]{Johansson:2011}
{Johansson}, D., {Sigurdarson}, H., \& {Horellou}, C. 2011, \aap, 527, A117

\bibitem[{{Kennicutt} \& {Evans}(2012)}]{Kennicutt:2012}
{Kennicutt}, R.~C. \& {Evans}, N.~J. 2012, \araa, 50, 531

\bibitem[{{Knudsen} {et~al.}(2008){Knudsen}, {van der Werf}, \&
  {Kneib}}]{Knudsen:2008}
{Knudsen}, K.~K., {van der Werf}, P.~P., \& {Kneib}, J.-P. 2008, \mnras, 384,
  1611

\bibitem[{{Lagache} {et~al.}(1999){Lagache}, {Abergel}, {Boulanger},
  {D{\'e}sert}, \& {Puget}}]{Lagache:1999}
{Lagache}, G., {Abergel}, A., {Boulanger}, F., {D{\'e}sert}, F.~X., \& {Puget},
  J.-L. 1999, \aap, 344, 322

\bibitem[{{MacGregor} {et~al.}(2013){MacGregor}, {Wilner}, {Rosenfeld},
  {Andrews}, {Matthews}, {Hughes}, {Booth}, {Chiang}, {Graham}, {Kalas},
  {Kennedy}, \& {Sibthorpe}}]{MacGregor:2013}
{MacGregor}, M.~A., {Wilner}, D.~J., {Rosenfeld}, K.~A., {et~al.} 2013, \apjl,
  762, L21

\bibitem[{{Maiolino} {et~al.}(2015){Maiolino}, {Carniani}, {Fontana},
  {Vallini}, {Pentericci}, {Ferrara}, {Vanzella}, {Grazian}, {Gallerani},
  {Castellano}, {Cristiani}, {Brammer}, {Santini}, {Wagg}, \&
  {Williams}}]{Maiolino:2015}
{Maiolino}, R., {Carniani}, S., {Fontana}, A., {et~al.} 2015, \mnras, 452, 54

\bibitem[{{Maiolino} {et~al.}(2008){Maiolino}, {Nagao}, {Grazian}, {Cocchia},
  {Marconi}, {Mannucci}, {Cimatti}, {Pipino}, {Ballero}, {Calura}, {Chiappini},
  {Fontana}, {Granato}, {Matteucci}, {Pastorini}, {Pentericci}, {Risaliti},
  {Salvati}, \& {Silva}}]{Maiolino:2008}
{Maiolino}, R., {Nagao}, T., {Grazian}, A., {et~al.} 2008, \aap, 488, 463

\bibitem[{{Moster} {et~al.}(2011){Moster}, {Somerville}, {Newman}, \&
  {Rix}}]{Moster:2011}
{Moster}, B.~P., {Somerville}, R.~S., {Newman}, J.~A., \& {Rix}, H.-W. 2011,
  \apj, 731, 113

\bibitem[{{Ono} {et~al.}(2014){Ono}, {Ouchi}, {Kurono}, \& {Momose}}]{Ono:2014}
{Ono}, Y., {Ouchi}, M., {Kurono}, Y., \& {Momose}, R. 2014, \apj, 795, 5

\bibitem[{{Ono} {et~al.}(2012){Ono}, {Ouchi}, {Mobasher}, {Dickinson},
  {Penner}, {Shimasaku}, {Weiner}, {Kartaltepe}, {Nakajima}, {Nayyeri},
  {Stern}, {Kashikawa}, \& {Spinrad}}]{Ono:2012}
{Ono}, Y., {Ouchi}, M., {Mobasher}, B., {et~al.} 2012, \apj, 744, 83

\bibitem[{{Ota} {et~al.}(2014){Ota}, {Walter}, {Ohta}, {Hatsukade}, {Carilli},
  {da Cunha}, {Gonz{\'a}lez-L{\'o}pez}, {Decarli}, {Hodge}, {Nagai}, {Egami},
  {Jiang}, {Iye}, {Kashikawa}, {Riechers}, {Bertoldi}, {Cox}, {Neri}, \&
  {Weiss}}]{Ota:2014}
{Ota}, K., {Walter}, F., {Ohta}, K., {et~al.} 2014, \apj, 792, 34

\bibitem[{{Ouchi} {et~al.}(2013){Ouchi}, {Ellis}, {Ono}, {Nakanishi}, {Kohno},
  {Momose}, {Kurono}, {Ashby}, {Shimasaku}, {Willner}, {Fazio}, {Tamura}, \&
  {Iono}}]{Ouchi:2013}
{Ouchi}, M., {Ellis}, R., {Ono}, Y., {et~al.} 2013, \apj, 778, 102

\bibitem[{{Puget} {et~al.}(1996){Puget}, {Abergel}, {Bernard}, {Boulanger},
  {Burton}, {Desert}, \& {Hartmann}}]{Puget:1996}
{Puget}, J.-L., {Abergel}, A., {Bernard}, J.-P., {et~al.} 1996, \aap, 308, L5

\bibitem[{{Rodighiero} {et~al.}(2011){Rodighiero}, {Daddi}, {Baronchelli},
  {Cimatti}, {Renzini}, {Aussel}, {Popesso}, {Lutz}, {Andreani}, {Berta},
  {Cava}, {Elbaz}, {Feltre}, {Fontana}, {F{\"o}rster Schreiber},
  {Franceschini}, {Genzel}, {Grazian}, {Gruppioni}, {Ilbert}, {Le Floch},
  {Magdis}, {Magliocchetti}, {Magnelli}, {Maiolino}, {McCracken}, {Nordon},
  {Poglitsch}, {Santini}, {Pozzi}, {Riguccini}, {Tacconi}, {Wuyts}, \&
  {Zamorani}}]{Rodighiero:2011}
{Rodighiero}, G., {Daddi}, E., {Baronchelli}, I., {et~al.} 2011, \apjl, 739,
  L40

\bibitem[{{Schmidt} {et~al.}(2015){Schmidt}, {M{\'e}nard}, {Scranton},
  {Morrison}, {Rahman}, \& {Hopkins}}]{Schmidt:2015}
{Schmidt}, S.~J., {M{\'e}nard}, B., {Scranton}, R., {et~al.} 2015, \mnras, 446,
  2696

\bibitem[{{Scott} {et~al.}(2012){Scott}, {Wilson}, {Aretxaga}, {Austermann},
  {Chapin}, {Dunlop}, {Ezawa}, {Halpern}, {Hatsukade}, {Hughes}, {Kawabe},
  {Kim}, {Kohno}, {Lowenthal}, {Monta{\~n}a}, {Nakanishi}, {Oshima}, {Sanders},
  {Scott}, {Scoville}, {Tamura}, {Welch}, {Yun}, \& {Zeballos}}]{Scott:2012}
{Scott}, K.~S., {Wilson}, G.~W., {Aretxaga}, I., {et~al.} 2012, \mnras, 423,
  575

\bibitem[{{Scott} {et~al.}(2010){Scott}, {Yun}, {Wilson}, {Austermann},
  {Aguilar}, {Aretxaga}, {Ezawa}, {Ferrusca}, {Hatsukade}, {Hughes}, {Iono},
  {Giavalisco}, {Kawabe}, {Kohno}, {Mauskopf}, {Oshima}, {Perera}, {Rand},
  {Tamura}, {Tosaki}, {Velazquez}, {Williams}, \& {Zeballos}}]{Scott:2010}
{Scott}, K.~S., {Yun}, M.~S., {Wilson}, G.~W., {et~al.} 2010, \mnras, 405, 2260

\bibitem[{{Shibuya} {et~al.}(2012){Shibuya}, {Kashikawa}, {Ota}, {Iye},
  {Ouchi}, {Furusawa}, {Shimasaku}, \& {Hattori}}]{Shibuya:2012a}
{Shibuya}, T., {Kashikawa}, N., {Ota}, K., {et~al.} 2012, \apj, 752, 114

\bibitem[{{Shimizu} {et~al.}(2012){Shimizu}, {Yoshida}, \&
  {Okamoto}}]{Shimizu:2012}
{Shimizu}, I., {Yoshida}, N., \& {Okamoto}, T. 2012, \mnras, 427, 2866

\bibitem[{{Simpson} {et~al.}(2014){Simpson}, {Swinbank}, {Smail}, {Alexander},
  {Brandt}, {Bertoldi}, {de Breuck}, {Chapman}, {Coppin}, {da Cunha},
  {Danielson}, {Dannerbauer}, {Greve}, {Hodge}, {Ivison}, {Karim}, {Knudsen},
  {Poggianti}, {Schinnerer}, {Thomson}, {Walter}, {Wardlow}, {Wei{\ss}}, \&
  {van der Werf}}]{simpson:2014}
{Simpson}, J.~M., {Swinbank}, A.~M., {Smail}, I., {et~al.} 2014, \apj, 788, 125

\bibitem[{{Toft} {et~al.}(2014){Toft}, {Smol{\v c}i{\'c}}, {Magnelli}, {Karim},
  {Zirm}, {Michalowski}, {Capak}, {Sheth}, {Schawinski}, {Krogager}, {Wuyts},
  {Sanders}, {Man}, {Lutz}, {Staguhn}, {Berta}, {Mccracken}, {Krpan}, \&
  {Riechers}}]{Toft:2014}
{Toft}, S., {Smol{\v c}i{\'c}}, V., {Magnelli}, B., {et~al.} 2014, \apj, 782,
  68

\bibitem[{{Vanzella} {et~al.}(2011){Vanzella}, {Pentericci}, {Fontana},
  {Grazian}, {Castellano}, {Boutsia}, {Cristiani}, {Dickinson}, {Gallozzi},
  {Giallongo}, {Giavalisco}, {Maiolino}, {Moorwood}, {Paris}, \&
  {Santini}}]{Vanzella:2011}
{Vanzella}, E., {Pentericci}, L., {Fontana}, A., {et~al.} 2011, \apjl, 730, L35

\bibitem[{{Viero} {et~al.}(2013){Viero}, {Moncelsi}, {Quadri}, {Arumugam},
  {Assef}, {B{\'e}thermin}, {Bock}, {Bridge}, {Casey}, {Conley}, {Cooray},
  {Farrah}, {Glenn}, {Heinis}, {Ibar}, {Ikarashi}, {Ivison}, {Kohno},
  {Marsden}, {Oliver}, {Roseboom}, {Schulz}, {Scott}, {Serra}, {Vaccari},
  {Vieira}, {Wang}, {Wardlow}, {Wilson}, {Yun}, \& {Zemcov}}]{Viero:2013a}
{Viero}, M.~P., {Moncelsi}, L., {Quadri}, R.~F., {et~al.} 2013, \apj, 779, 32

\bibitem[{{Wei{\ss}} {et~al.}(2013){Wei{\ss}}, {De Breuck}, {Marrone},
  {Vieira}, {Aguirre}, {Aird}, {Aravena}, {Ashby}, {Bayliss}, {Benson},
  {B{\'e}thermin}, {Biggs}, {Bleem}, {Bock}, {Bothwell}, {Bradford}, {Brodwin},
  {Carlstrom}, {Chang}, {Chapman}, {Crawford}, {Crites}, {de Haan}, {Dobbs},
  {Downes}, {Fassnacht}, {George}, {Gladders}, {Gonzalez}, {Greve},
  {Halverson}, {Hezaveh}, {High}, {Holder}, {Holzapfel}, {Hoover}, {Hrubes},
  {Husband}, {Keisler}, {Lee}, {Leitch}, {Lueker}, {Luong-Van}, {Malkan},
  {McIntyre}, {McMahon}, {Mehl}, {Menten}, {Meyer}, {Murphy}, {Padin},
  {Plagge}, {Reichardt}, {Rest}, {Rosenman}, {Ruel}, {Ruhl}, {Schaffer},
  {Shirokoff}, {Spilker}, {Stalder}, {Staniszewski}, {Stark}, {Story},
  {Vanderlinde}, {Welikala}, \& {Williamson}}]{Weis:2013}
{Wei{\ss}}, A., {De Breuck}, C., {Marrone}, D.~P., {et~al.} 2013, \apj, 767, 88

\bibitem[{{Wei{\ss}} {et~al.}(2009){Wei{\ss}}, {Kov{\'a}cs}, {Coppin}, {Greve},
  {Walter}, {Smail}, {Dunlop}, {Knudsen}, {Alexander}, {Bertoldi}, {Brandt},
  {Chapman}, {Cox}, {Dannerbauer}, {De Breuck}, {Gawiser}, {Ivison}, {Lutz},
  {Menten}, {Koekemoer}, {Kreysa}, {Kurczynski}, {Rix}, {Schinnerer}, \& {van
  der Werf}}]{Weis:2009}
{Wei{\ss}}, A., {Kov{\'a}cs}, A., {Coppin}, K., {et~al.} 2009, \apj, 707, 1201

\bibitem[{{Willott} {et~al.}(2013){Willott}, {Omont}, \&
  {Bergeron}}]{Willott:2013}
{Willott}, C.~J., {Omont}, A., \& {Bergeron}, J. 2013, \apj, 770, 13

\bibitem[{{Yun} {et~al.}(2012){Yun}, {Scott}, {Guo}, {Aretxaga}, {Giavalisco},
  {Austermann}, {Capak}, {Chen}, {Ezawa}, {Hatsukade}, {Hughes}, {Iono},
  {Johnson}, {Kawabe}, {Kohno}, {Lowenthal}, {Miller}, {Morrison}, {Oshima},
  {Perera}, {Salvato}, {Silverman}, {Tamura}, {Williams}, \&
  {Wilson}}]{Yun:2012}
{Yun}, M.~S., {Scott}, K.~S., {Guo}, Y., {et~al.} 2012, \mnras, 420, 957

\end{thebibliography}


\begin{thebibliography}{38}
\expandafter\ifx\csname natexlab\endcsname\relax\def\natexlab#1{#1}\fi

\bibitem[{{Blain} {et~al.}(1999){Blain}, {Moller}, \& {Maller}}]{Blain:1999}
{Blain}, A.~W., {Moller}, O., \& {Maller}, A.~H. 1999, \mnras, 303, 423

\bibitem[{{Blain} {et~al.}(2002){Blain}, {Smail}, {Ivison}, {Kneib}, \&
  {Frayer}}]{Blain:2002}
{Blain}, A.~W., {Smail}, I., {Ivison}, R.~J., {Kneib}, J.-P., \& {Frayer},
  D.~T. 2002, \physrep, 369, 111

\bibitem[{{Cai} {et~al.}(2013){Cai}, {Lapi}, {Xia}, {De Zotti}, {Negrello},
  {Gruppioni}, {Rigby}, {Castex}, {Delabrouille}, \& {Danese}}]{Cai:2013}
{Cai}, Z.-Y., {Lapi}, A., {Xia}, J.-Q., {et~al.} 2013, \apj, 768, 21

\bibitem[{{Casey} {et~al.}(2013){Casey}, {Chen}, {Cowie}, {Barger}, {Capak},
  {Ilbert}, {Koss}, {Lee}, {Le Floc'h}, {Sanders}, \& {Williams}}]{Casey:2013}
{Casey}, C.~M., {Chen}, C.-C., {Cowie}, L.~L., {et~al.} 2013, \mnras, 436, 1919

\bibitem[{{Casey} {et~al.}(2014){Casey}, {Narayanan}, \&
  {Cooray}}]{Casey:2014a}
{Casey}, C.~M., {Narayanan}, D., \& {Cooray}, A. 2014, \physrep, 541, 45

\bibitem[{{Coppin} {et~al.}(2006){Coppin}, {Chapin}, {Mortier}, {Scott},
  {Borys}, {Dunlop}, {Halpern}, {Hughes}, {Pope}, {Scott}, {Serjeant}, {Wagg},
  {Alexander}, {Almaini}, {Aretxaga}, {Babbedge}, {Best}, {Blain}, {Chapman},
  {Clements}, {Crawford}, {Dunne}, {Eales}, {Edge}, {Farrah}, {Gazta{\~n}aga},
  {Gear}, {Granato}, {Greve}, {Fox}, {Ivison}, {Jarvis}, {Jenness}, {Lacey},
  {Lepage}, {Mann}, {Marsden}, {Martinez-Sansigre}, {Oliver}, {Page},
  {Peacock}, {Pearson}, {Percival}, {Priddey}, {Rawlings}, {Rowan-Robinson},
  {Savage}, {Seigar}, {Sekiguchi}, {Silva}, {Simpson}, {Smail}, {Stevens},
  {Takagi}, {Vaccari}, {van Kampen}, \& {Willott}}]{Coppin:2006}
{Coppin}, K., {Chapin}, E.~L., {Mortier}, A.~M.~J., {et~al.} 2006, \mnras, 372,
  1621

\bibitem[{{Dayal} {et~al.}(2014){Dayal}, {Ferrara}, {Dunlop}, \&
  {Pacucci}}]{Dayal:2014}
{Dayal}, P., {Ferrara}, A., {Dunlop}, J.~S., \& {Pacucci}, F. 2014, \mnras,
  445, 2545

\bibitem[{{Eales} {et~al.}(1999){Eales}, {Lilly}, {Gear}, {Dunne}, {Bond},
  {Hammer}, {Le F{\`e}vre}, \& {Crampton}}]{Eales:1999}
{Eales}, S., {Lilly}, S., {Gear}, W., {et~al.} 1999, \apj, 515, 518

\bibitem[{{Fixsen} {et~al.}(1998){Fixsen}, {Dwek}, {Mather}, {Bennett}, \&
  {Shafer}}]{Fixsen:1998}
{Fixsen}, D.~J., {Dwek}, E., {Mather}, J.~C., {Bennett}, C.~L., \& {Shafer},
  R.~A. 1998, \apj, 508, 123

\bibitem[{{Gehrels}(1986)}]{Gehrels:1986}
{Gehrels}, N. 1986, \apj, 303, 336

\bibitem[{{Greve} {et~al.}(2012){Greve}, {Vieira}, {Wei{\ss}}, {Aguirre},
  {Aird}, {Ashby}, {Benson}, {Bleem}, {Bradford}, {Brodwin}, {Carlstrom},
  {Chang}, {Chapman}, {Crawford}, {de Breuck}, {de Haan}, {Dobbs}, {Downes},
  {Fassnacht}, {Fazio}, {George}, {Gladders}, {Gonzalez}, {Halverson},
  {Hezaveh}, {High}, {Holder}, {Holzapfel}, {Hoover}, {Hrubes}, {Johnson},
  {Keisler}, {Knox}, {Lee}, {Leitch}, {Lueker}, {Luong-Van}, {Malkan},
  {Marrone}, {McIntyre}, {McMahon}, {Mehl}, {Menten}, {Meyer}, {Montroy},
  {Murphy}, {Natoli}, {Padin}, {Plagge}, {Pryke}, {Reichardt}, {Rest},
  {Rosenman}, {Ruel}, {Ruhl}, {Schaffer}, {Sharon}, {Shaw}, {Shirokoff},
  {Stalder}, {Stanford}, {Staniszewski}, {Stark}, {Story}, {Vanderlinde},
  {Walsh}, {Welikala}, \& {Williamson}}]{Greve:2012}
{Greve}, T.~R., {Vieira}, J.~D., {Wei{\ss}}, A., {et~al.} 2012, \apj, 756, 101

\bibitem[{{Hatsukade} {et~al.}(2013){Hatsukade}, {Ohta}, {Seko}, {Yabe}, \&
  {Akiyama}}]{Hatsukade:2013}
{Hatsukade}, B., {Ohta}, K., {Seko}, A., {Yabe}, K., \& {Akiyama}, M. 2013,
  \apjl, 769, L27

\bibitem[{{Hayward} {et~al.}(2013){Hayward}, {Narayanan}, {Kere{\v s}},
  {Jonsson}, {Hopkins}, {Cox}, \& {Hernquist}}]{Hayward:2013}
{Hayward}, C.~C., {Narayanan}, D., {Kere{\v s}}, D., {et~al.} 2013, \mnras,
  428, 2529

\bibitem[{{Hogg} \& {Turner}(1998)}]{Hogg:1998}
{Hogg}, D.~W. \& {Turner}, E.~L. 1998, \pasp, 110, 727

\bibitem[{{Hughes} {et~al.}(1998){Hughes}, {Serjeant}, {Dunlop},
  {Rowan-Robinson}, {Blain}, {Mann}, {Ivison}, {Peacock}, {Efstathiou}, {Gear},
  {Oliver}, {Lawrence}, {Longair}, {Goldschmidt}, \& {Jenness}}]{Hughes:1998}
{Hughes}, D.~H., {Serjeant}, S., {Dunlop}, J., {et~al.} 1998, \nat, 394, 241

\bibitem[{{Kashlinsky} {et~al.}(2005){Kashlinsky}, {Arendt}, {Mather}, \&
  {Moseley}}]{Kashlinsky:2005a}
{Kashlinsky}, A., {Arendt}, R.~G., {Mather}, J., \& {Moseley}, S.~H. 2005,
  \nat, 438, 45

\bibitem[{{Kennicutt} \& {Evans}(2012)}]{Kennicutt:2012}
{Kennicutt}, R.~C. \& {Evans}, N.~J. 2012, \araa, 50, 531

\bibitem[{{Lagache} {et~al.}(1999){Lagache}, {Abergel}, {Boulanger},
  {D{\'e}sert}, \& {Puget}}]{Lagache:1999}
{Lagache}, G., {Abergel}, A., {Boulanger}, F., {D{\'e}sert}, F.~X., \& {Puget},
  J.-L. 1999, \aap, 344, 322

\bibitem[{{MacGregor} {et~al.}(2013){MacGregor}, {Wilner}, {Rosenfeld},
  {Andrews}, {Matthews}, {Hughes}, {Booth}, {Chiang}, {Graham}, {Kalas},
  {Kennedy}, \& {Sibthorpe}}]{MacGregor:2013}
{MacGregor}, M.~A., {Wilner}, D.~J., {Rosenfeld}, K.~A., {et~al.} 2013, \apjl,
  762, L21

\bibitem[{{Maiolino} {et~al.}(2008){Maiolino}, {Nagao}, {Grazian}, {Cocchia},
  {Marconi}, {Mannucci}, {Cimatti}, {Pipino}, {Ballero}, {Calura}, {Chiappini},
  {Fontana}, {Granato}, {Matteucci}, {Pastorini}, {Pentericci}, {Risaliti},
  {Salvati}, \& {Silva}}]{Maiolino:2008}
{Maiolino}, R., {Nagao}, T., {Grazian}, A., {et~al.} 2008, \aap, 488, 463

\bibitem[{{Ono} {et~al.}(2014){Ono}, {Ouchi}, {Kurono}, \& {Momose}}]{Ono:2014}
{Ono}, Y., {Ouchi}, M., {Kurono}, Y., \& {Momose}, R. 2014, \apj, 795, 5

\bibitem[{{Ono} {et~al.}(2012){Ono}, {Ouchi}, {Mobasher}, {Dickinson},
  {Penner}, {Shimasaku}, {Weiner}, {Kartaltepe}, {Nakajima}, {Nayyeri},
  {Stern}, {Kashikawa}, \& {Spinrad}}]{Ono:2012}
{Ono}, Y., {Ouchi}, M., {Mobasher}, B., {et~al.} 2012, \apj, 744, 83

\bibitem[{{Ota} {et~al.}(2014){Ota}, {Walter}, {Ohta}, {Hatsukade}, {Carilli},
  {da Cunha}, {Gonz{\'a}lez-L{\'o}pez}, {Decarli}, {Hodge}, {Nagai}, {Egami},
  {Jiang}, {Iye}, {Kashikawa}, {Riechers}, {Bertoldi}, {Cox}, {Neri}, \&
  {Weiss}}]{Ota:2014}
{Ota}, K., {Walter}, F., {Ohta}, K., {et~al.} 2014, \apj, 792, 34

\bibitem[{{Ouchi} {et~al.}(2013){Ouchi}, {Ellis}, {Ono}, {Nakanishi}, {Kohno},
  {Momose}, {Kurono}, {Ashby}, {Shimasaku}, {Willner}, {Fazio}, {Tamura}, \&
  {Iono}}]{Ouchi:2013}
{Ouchi}, M., {Ellis}, R., {Ono}, Y., {et~al.} 2013, \apj, 778, 102

\bibitem[{{Puget} {et~al.}(1996){Puget}, {Abergel}, {Bernard}, {Boulanger},
  {Burton}, {Desert}, \& {Hartmann}}]{Puget:1996}
{Puget}, J.-L., {Abergel}, A., {Bernard}, J.-P., {et~al.} 1996, \aap, 308, L5

\bibitem[{{Rodighiero} {et~al.}(2011){Rodighiero}, {Daddi}, {Baronchelli},
  {Cimatti}, {Renzini}, {Aussel}, {Popesso}, {Lutz}, {Andreani}, {Berta},
  {Cava}, {Elbaz}, {Feltre}, {Fontana}, {F{\"o}rster Schreiber},
  {Franceschini}, {Genzel}, {Grazian}, {Gruppioni}, {Ilbert}, {Le Floch},
  {Magdis}, {Magliocchetti}, {Magnelli}, {Maiolino}, {McCracken}, {Nordon},
  {Poglitsch}, {Santini}, {Pozzi}, {Riguccini}, {Tacconi}, {Wuyts}, \&
  {Zamorani}}]{Rodighiero:2011}
{Rodighiero}, G., {Daddi}, E., {Baronchelli}, I., {et~al.} 2011, \apjl, 739,
  L40

\bibitem[{{Schmidt} {et~al.}(2015){Schmidt}, {M{\'e}nard}, {Scranton},
  {Morrison}, {Rahman}, \& {Hopkins}}]{Schmidt:2015}
{Schmidt}, S.~J., {M{\'e}nard}, B., {Scranton}, R., {et~al.} 2015, \mnras, 446,
  2696

\bibitem[{{Scott} {et~al.}(2012){Scott}, {Wilson}, {Aretxaga}, {Austermann},
  {Chapin}, {Dunlop}, {Ezawa}, {Halpern}, {Hatsukade}, {Hughes}, {Kawabe},
  {Kim}, {Kohno}, {Lowenthal}, {Monta{\~n}a}, {Nakanishi}, {Oshima}, {Sanders},
  {Scott}, {Scoville}, {Tamura}, {Welch}, {Yun}, \& {Zeballos}}]{Scott:2012}
{Scott}, K.~S., {Wilson}, G.~W., {Aretxaga}, I., {et~al.} 2012, \mnras, 423,
  575

\bibitem[{{Shibuya} {et~al.}(2012){Shibuya}, {Kashikawa}, {Ota}, {Iye},
  {Ouchi}, {Furusawa}, {Shimasaku}, \& {Hattori}}]{Shibuya:2012a}
{Shibuya}, T., {Kashikawa}, N., {Ota}, K., {et~al.} 2012, \apj, 752, 114

\bibitem[{{Shimizu} {et~al.}(2012){Shimizu}, {Yoshida}, \&
  {Okamoto}}]{Shimizu:2012}
{Shimizu}, I., {Yoshida}, N., \& {Okamoto}, T. 2012, \mnras, 427, 2866

\bibitem[{{Simpson} {et~al.}(2014){Simpson}, {Swinbank}, {Smail}, {Alexander},
  {Brandt}, {Bertoldi}, {de Breuck}, {Chapman}, {Coppin}, {da Cunha},
  {Danielson}, {Dannerbauer}, {Greve}, {Hodge}, {Ivison}, {Karim}, {Knudsen},
  {Poggianti}, {Schinnerer}, {Thomson}, {Walter}, {Wardlow}, {Wei{\ss}}, \&
  {van der Werf}}]{simpson:2014}
{Simpson}, J.~M., {Swinbank}, A.~M., {Smail}, I., {et~al.} 2014, \apj, 788, 125

\bibitem[{{Smail} {et~al.}(2002){Smail}, {Ivison}, {Blain}, \&
  {Kneib}}]{Smail:2002}
{Smail}, I., {Ivison}, R.~J., {Blain}, A.~W., \& {Kneib}, J.-P. 2002, \mnras,
  331, 495

\bibitem[{{Toft} {et~al.}(2014){Toft}, {Smol{\v c}i{\'c}}, {Magnelli}, {Karim},
  {Zirm}, {Michalowski}, {Capak}, {Sheth}, {Schawinski}, {Krogager}, {Wuyts},
  {Sanders}, {Man}, {Lutz}, {Staguhn}, {Berta}, {Mccracken}, {Krpan}, \&
  {Riechers}}]{Toft:2014}
{Toft}, S., {Smol{\v c}i{\'c}}, V., {Magnelli}, B., {et~al.} 2014, \apj, 782,
  68

\bibitem[{{Vanzella} {et~al.}(2011){Vanzella}, {Pentericci}, {Fontana},
  {Grazian}, {Castellano}, {Boutsia}, {Cristiani}, {Dickinson}, {Gallozzi},
  {Giallongo}, {Giavalisco}, {Maiolino}, {Moorwood}, {Paris}, \&
  {Santini}}]{Vanzella:2011}
{Vanzella}, E., {Pentericci}, L., {Fontana}, A., {et~al.} 2011, \apjl, 730, L35

\bibitem[{{Viero} {et~al.}(2013){Viero}, {Moncelsi}, {Quadri}, {Arumugam},
  {Assef}, {B{\'e}thermin}, {Bock}, {Bridge}, {Casey}, {Conley}, {Cooray},
  {Farrah}, {Glenn}, {Heinis}, {Ibar}, {Ikarashi}, {Ivison}, {Kohno},
  {Marsden}, {Oliver}, {Roseboom}, {Schulz}, {Scott}, {Serra}, {Vaccari},
  {Vieira}, {Wang}, {Wardlow}, {Wilson}, {Yun}, \& {Zemcov}}]{Viero:2013a}
{Viero}, M.~P., {Moncelsi}, L., {Quadri}, R.~F., {et~al.} 2013, \apj, 779, 32

\bibitem[{{Wardlow} {et~al.}(2011){Wardlow}, {Smail}, {Coppin}, {Alexander},
  {Brandt}, {Danielson}, {Luo}, {Swinbank}, {Walter}, {Wei{\ss}}, {Xue},
  {Zibetti}, {Bertoldi}, {Biggs}, {Chapman}, {Dannerbauer}, {Dunlop},
  {Gawiser}, {Ivison}, {Knudsen}, {Kov{\'a}cs}, {Lacey}, {Menten}, {Padilla},
  {Rix}, \& {van der Werf}}]{Wardlow:2011}
{Wardlow}, J.~L., {Smail}, I., {Coppin}, K.~E.~K., {et~al.} 2011, \mnras, 415,
  1479

\bibitem[{{Wei{\ss}} {et~al.}(2009){Wei{\ss}}, {Kov{\'a}cs}, {Coppin}, {Greve},
  {Walter}, {Smail}, {Dunlop}, {Knudsen}, {Alexander}, {Bertoldi}, {Brandt},
  {Chapman}, {Cox}, {Dannerbauer}, {De Breuck}, {Gawiser}, {Ivison}, {Lutz},
  {Menten}, {Koekemoer}, {Kreysa}, {Kurczynski}, {Rix}, {Schinnerer}, \& {van
  der Werf}}]{Weis:2009}
{Wei{\ss}}, A., {Kov{\'a}cs}, A., {Coppin}, K., {et~al.} 2009, \apj, 707, 1201

\bibitem[{{Willott} {et~al.}(2013){Willott}, {Omont}, \&
  {Bergeron}}]{Willott:2013}
{Willott}, C.~J., {Omont}, A., \& {Bergeron}, J. 2013, \apj, 770, 13

\end{thebibliography}
\appendix

\section{ALMA noise fluctuations}\label{sec:appendix_A}
 
In Sec. \ref{sec:diff}, we have defined $f_c$ as the ratio between negative and
positive detections and we have estimated how $f_c$ depends on the S/N of our
observations. Since this study shows that some of the real sources can be
spurious because of noise fluctuations, we further verify the
reliability of our catalogue in this Appendix  by estimating the $f_c$  value expected in blank
fields observed with ALMA. In a pure blank field, the positive and negative
sources are only caused by noise fluctuation, so we expect $f_c \sim 1$ at any
S/N level. 


We used the simobserve CASA v.4.2.1  task to produce synthetic
interferometric observations of a blank field placed at the RA = 22:28:12.28
and DEC = -35:089:59.6, which are the coordinates of the deepest continuum map
(Field $a$ Table \ref{tab:field}). As for the CASA input, we required that antenna configuration would be the same as those used during the observations. Furthermore, we added a thermal noise component  by setting the parameter thermal noise to \emph{tsys-atm} with a precipitable water vapour of 1.1 mm and ambient temperature of 269 K, which are typical values of our observations. We simulated 300 continuum maps changing each time the parameter seed with a random value, which allows us to generate a random thermal noise for each observation.

\begin{figure}
   \centering 
   \includegraphics[width=.5\textwidth]{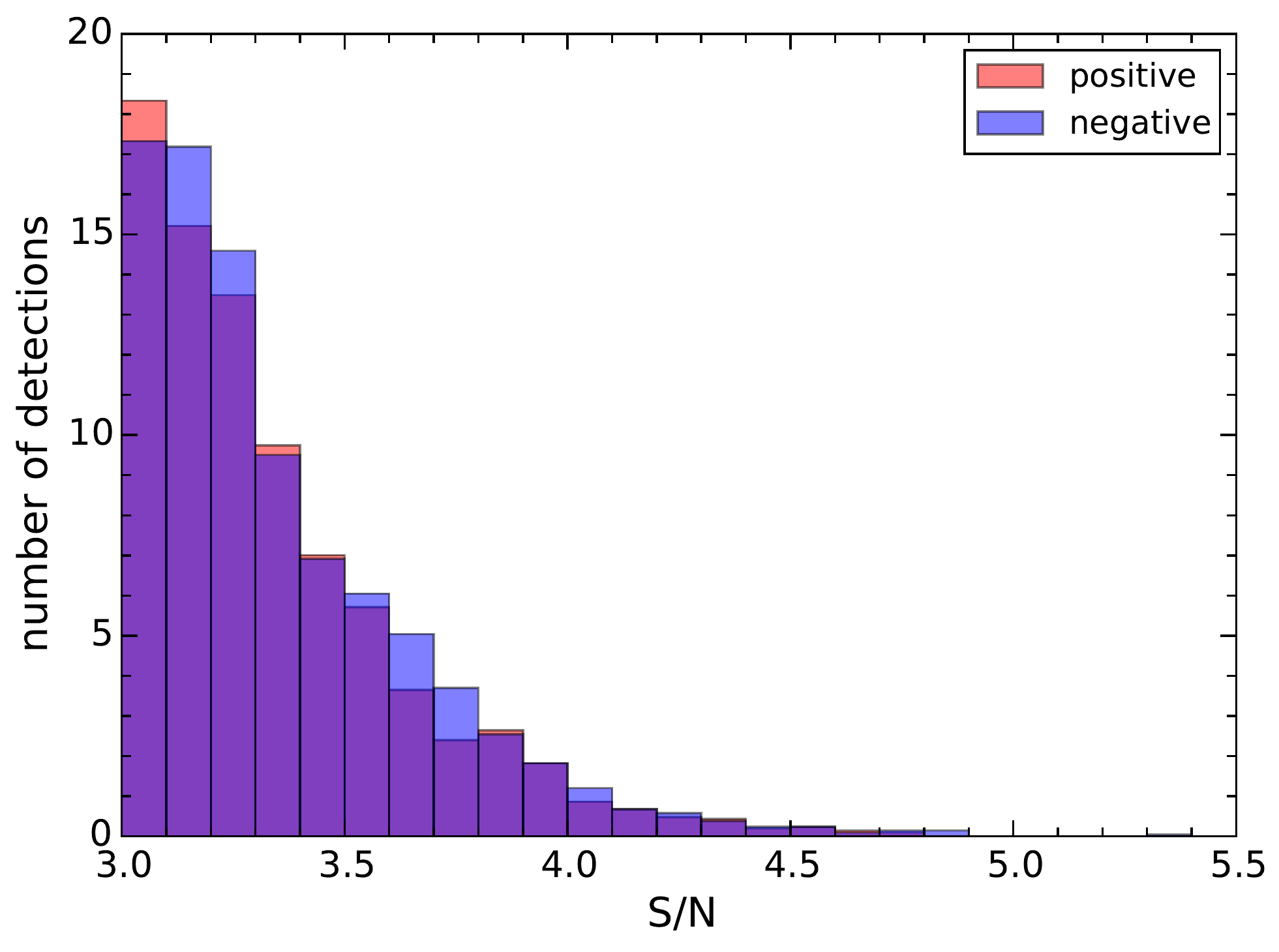}
   \includegraphics[width=.5\textwidth]{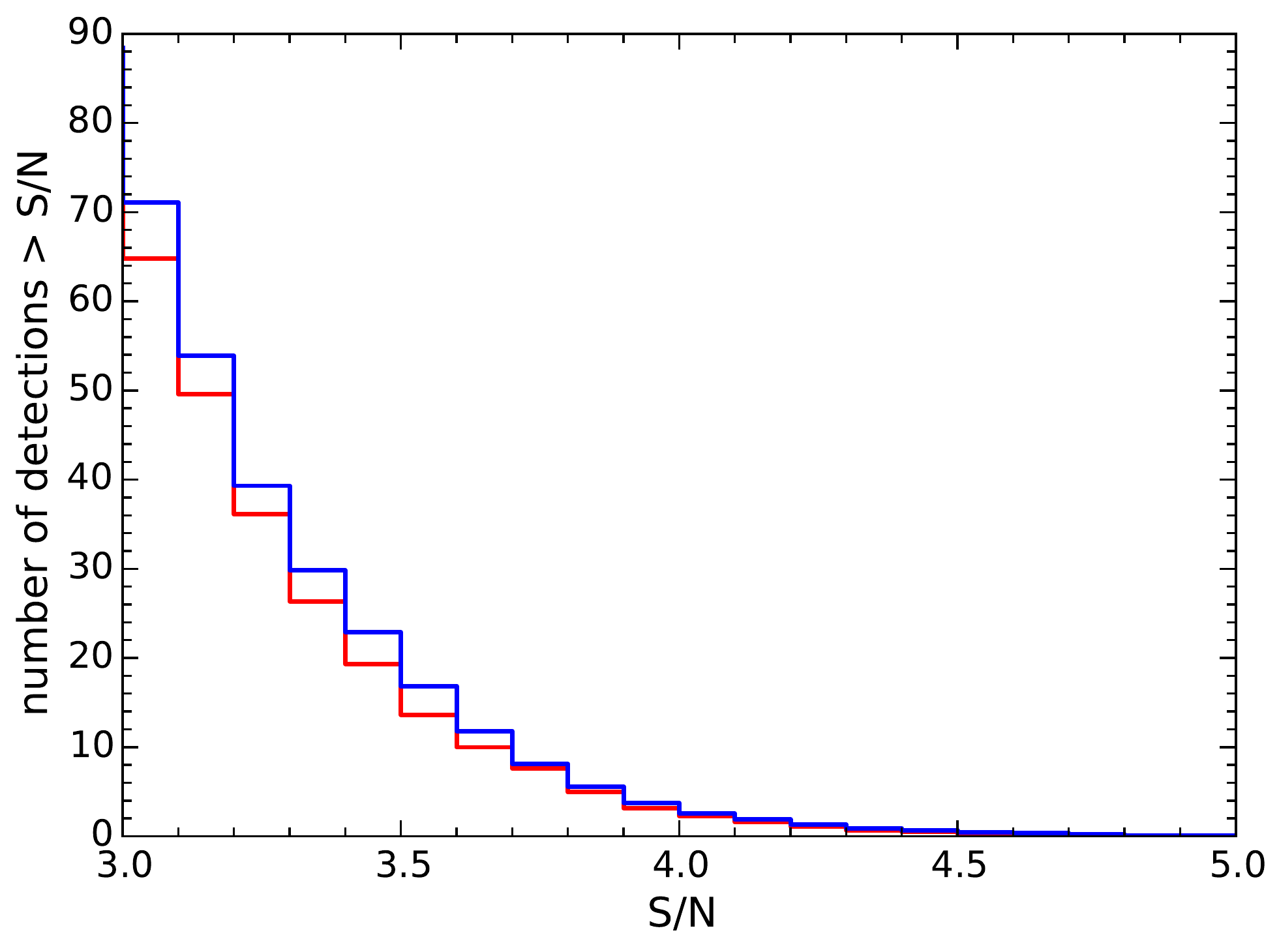}

\caption{\emph{Top:} number of positive (red) and negative (blue) sources detected in 300 continuum pure noise maps and normalised  for 18 continuum maps. \emph{Bottom:} cumulative distribution of positive (red) and negative (blue) detections.}
 \label{fig:histo_simulation}
   \end{figure}

We then applied the source extraction technique, mentioned in Section
\ref{sec:source_extraction}, on each mock continuum map. Figure
\ref{fig:histo_simulation} shows the number of positive and negative sources as
a function of S/N normalised to 18 continuum fields. The number of negative
sources is equal to those of positive ones at any S/N, indicating that the number of positive and negative spurious sources due to noise fluctuations is equal ($f_c \sim 1$ for each S/N). 

Since the number of spurious positive sources is almost equal to negative ones in a blank field, most of the positive sources detected in our observations with S/N$>3$ (Fig. \ref{fig:histo}) are likely to be real. 

\section{Flux error}\label{sec:appendix_B}
\begin{figure}
   \centering 
   \includegraphics[width=.5\textwidth]{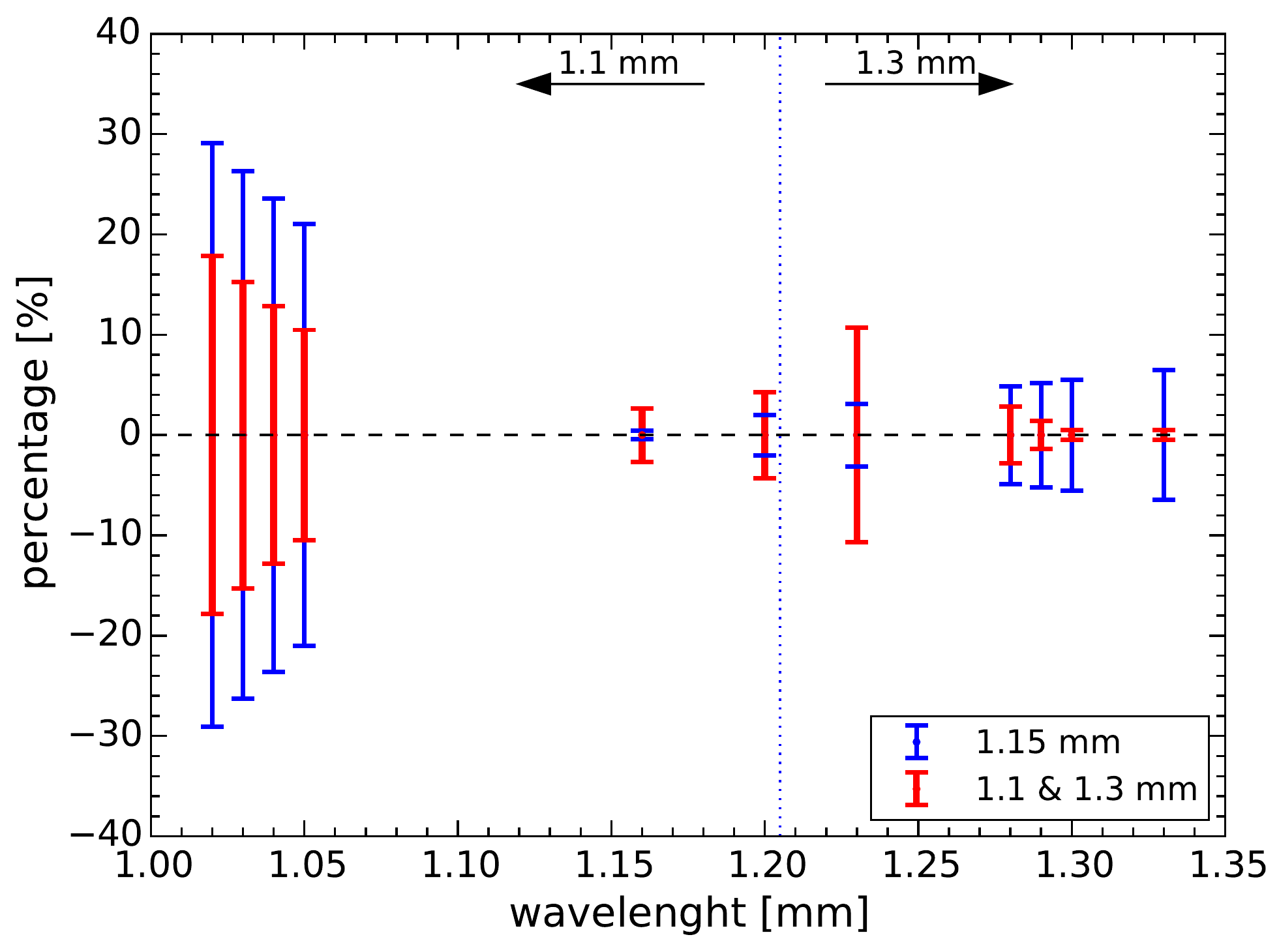}
        \label{fig:flux_error}
\caption{Flux error at different wavelength. The red error bars show the flux error  scaling the flux density  of the sources observed at $\lambda \leq 1.2$ mm to the flux density at 1.1 mm and those at $\lambda > 1.2$ mm to the flux density at 1.3 mm.  The blue error bars indicate the flux error combining all observations to 1.115 mm. }
   \end{figure}

None of the detected sources in this work has a spectroscopic redshift,
which prevents us from determining their SED or their flux densities at
different wavelengths. In section \ref{sec:diff}, we scaled the flux density of
the sources observed at $\lambda <$ 1.2 mm to the flux density at 1.1 mm, and
the sources observed at $\lambda >$ 1.2 mm are scaled to the flux density at
1.3 mm, by assuming a SED given by a greybody
with the following properties: $z = 2$, T = 35K, $\beta$ = 2, where T and
$\beta$ are dust temperature and dust emissivity index ($\varepsilon \propto
\lambda ^{-\beta}$), respectively. However, we are aware of the fact that only one photometric value for each source is not enough to constrain the properties of its SED. In this Appendix, we estimate how the assumed SED properties affect the outcomes of the flux-scaling procedure. To this aim, we vary the SED properties in the following ranges: $1< z< 6$, 20 $<$T $<$60 K, $1.5< \ \beta< 2$.
The errors are estimated as the maximum scatter obtained by varying  these
parameters with respect to the typical SED used in our observations.  
Figure \ref{fig:flux_error} shows  the flux error (red error bars) associated
with each continuum map resulting from scaling the flux density  of the sources
observed at $\lambda \leq 1.2$ mm to the flux density at 1.1 mm, and those at
$\lambda > 1.2$ mm to the flux density at 1.3 mm. At 1.3 mm, the flux errors
are smaller than those at 1.1 mm since the wavelength range of observations is
smaller ($\Delta\lambda \sim$ 0.15 mm) than  at 1.1 mm ($\Delta\lambda \sim$
0.20 mm).  The blue error bars show the flux error resulting from scaling all
observations to a common average wavelength of 1.15 mm. The latter show
that by rescaling all our ALMA observations to a single common wavelength there
is, in most cases, a significant increase of the flux errors. Indeed,  the
flux errors approach 30\% at
1.1 mm, while at 1.3 mm the flux errors are at least twice
as large as those resulting from splitting the number counts in two different
wavelength ranges. Since we aim at minimising the flux errors as much as
possible (and keeping them lower than the size of our flux bins), we split the
number counts into two different wavelength ranges so as to reduce the flux errors associated to each detected source, at the sacrifice of having slightly worse statistics.

\end{document}